\documentclass[aps,twocolumn,amsmath,amsfonts,amssymb,groupedaddress,nofootinbib,nobalancelastpage,floatfix,superscriptaddress,secnumarabic,preprintnumbers]{revtex4-2}
\pdfoutput=1
\usepackage{epsfig, graphicx, hyperref, slashed,xspace}
\usepackage[dvipsnames]{xcolor}
\usepackage[utf8]{inputenc}
\usepackage[T1]{fontenc}
\usepackage{cleveref}
\usepackage{soul}

\hypersetup{colorlinks=true,linkcolor=Maroon,citecolor=ForestGreen,filecolor=ForestGreen,urlcolor=ForestGreen}

\DeclareUnicodeCharacter{2212}{-}

\def\eg{{\it e.g.}}

\def\lag{{\cal L}}
\def\sss{\scriptscriptstyle}
\def\alphas{\alpha_{\sss s}}
\def\uR{u_R}
\def\uRb{\bar{u}_R}
\newcommand{\sigmav}{\langle \sigma v \rangle} 

\newcommand{\Li}{{\rm Li}_2}

\newcommand{\be}{\begin{equation}}
\newcommand{\ee}{\end{equation}}
\def\bsp#1\esp{\begin{split}#1\end{split}}
\def\bpm{\begin{pmatrix}}
\def\epm{\end{pmatrix}}

\newcommand{\ch}{{\sc CalcHEP}\xspace}
\newcommand{\fa}{{\sc FeynArts}\xspace}
\newcommand{\fr}{{\sc FeynRules}\xspace}
\newcommand{\ma}{{\sc MadAnalysis}~5\xspace}
\newcommand{\maddm}{{\sc MadDM}\xspace}
\newcommand{\mg}{{\sc MG5\_aMC}\xspace}
\newcommand{\micromegas}{{\sc micrOMEGAs}\xspace}
\newcommand{\nloct}{{\sc NLOCT}\xspace}
\def\dmsimpt{{\tt DMSimpt}\xspace}

\definecolor{coral}{RGB}{255,127,80}

\definecolor{benj}{rgb}{0.55, 0.0, 0.55}

\definecolor{jansgreen}{RGB}{60,180,20}

\usepackage[normalem]{ulem}

\newcommand{\SM}{{\tt S3M\_uR}\xspace}
\newcommand{\SD}{{\tt S3D\_uR}\xspace}
\newcommand{\FS}{{\tt F3S\_uR}\xspace}
\newcommand{\FC}{{\tt F3C\_uR}\xspace}
\newcommand{\FV}{{\tt F3V\_uR}\xspace}
\newcommand{\FW}{{\tt F3W\_uR}\xspace}
\newcommand{\FX}{{\tt F3X\_uR}\xspace}

\makeatletter\def\l@subsubsection#1#2{}\makeatother

\begin{document}

\preprint{TTK-23-19}

\title{Comprehensive exploration of t-channel simplified models of dark matter}

\author{Chiara Arina}
\affiliation{Centre for Cosmology, Particle Physics and Phenomenology (CP3), Universit\'e catholique de Louvain, B-1348 Louvain-la-Neuve, Belgium}

\author{Benjamin Fuks}
\affiliation{Laboratoire de Physique Th\'eorique et Hautes \'Energies (LPTHE), UMR 7589, Sorbonne Universit\'e et CNRS, 4 place Jussieu, 75252 Paris Cedex 05, France}

\author{Jan Heisig}
\affiliation{Institute for Theoretical Particle Physics and Cosmology, RWTH Aachen University, D-52056 Aachen, Germany}
\affiliation{Department of Physics, University of Virginia, Charlottesville, Virginia 22904-4714, USA}

\author{Michael Kr\"{a}mer}
\affiliation{Institute for Theoretical Particle Physics and Cosmology, RWTH Aachen University, D-52056 Aachen, Germany}

\author{Luca Mantani}
\affiliation{DAMTP, University of Cambridge, Wilberforce Road,
Cambridge, CB3 0WA, United Kingdom}

\author{Luca Panizzi}
\affiliation{Dipartimento di Fisica, Universit\`a della Calabria, I-87036 Arcavacata di Rende, Cosenza, Italy}
\affiliation{INFN-Cosenza, I-87036 Arcavacata di Rende, Cosenza, Italy}
\affiliation{School of Physics and Astronomy, University of Southampton, Highfield, Southampton SO17 1BJ, UK}

\begin{abstract}
    We analyse six classes of $t$-channel dark matter simplified models in which the Standard Model field content is extended by a coloured mediator and a dark matter state. The two new states are enforced to be odd under a new parity, while all Standard Model fields are taken even so that dark matter stability is guaranteed. We study several possibilities for the spin of the new particles and the self-conjugate property of the dark matter, and we focus on model configurations in which the dark matter couples to the right-handed up quark for simplicity. We investigate how the parameter spaces of the six models can be constrained by current and future cosmological, astrophysical and collider searches, and we highlight the strong complementary between those probes. Our results demonstrate that scenarios featuring a complex (non self-conjugate) dark matter field are excluded by cosmology and astrophysics alone, the only possibility to avoid these bounds being to invoke very weak couplings and mechanisms such as conversion-driven freeze-out. For models with self-conjugate dark matter, mediator and dark matter masses are pushed deep into the TeV regime, with the lower limits on the mediator mass reaching 3 to 4~TeV and those on the dark matter mass 1 to 2~TeV. In large parts of the parameter space these strong bounds are driven by same-sign mediator pair production, a channel so far not considered in the experimental analyses embedding $t$-channel dark matter model interpretations.
\end{abstract}

\maketitle


\section{Introduction}\label{sec:intro}

Despite convincing indirect evidence for dark matter (DM) in the Universe~\cite{Bertone:2010zza, Bertone:2016nfn}, its origin remains one of the main puzzling issues in particle physics, astrophysics and cosmology. A plethora of models have consequently been proposed to address this problem, many of these models assuming that DM interacts with the Standard Model (SM) in one way or the other. They all predict the existence of new particles and phenomena beyond the SM of particle physics, and they offer various handles to search for DM experimentally. However, direct searches in nuclear and electronic recoil experiments, indirect probes through the analysis of cosmic- and gamma-ray spectra, and the hunt for missing energy signals at particle colliders all returned negative results so far. As a consequence, limits have been set on many DM models, which all get more and more severely constrained. These bounds are generally explored either in a model-specific approach, or in a more general phenomenological-driven approach based on simplified models~\cite{Alwall:2008ag, LHCNewPhysicsWorkingGroup:2011mji} representing large classes of theories beyond the SM and covering a broad set of signatures. 

In such simplified models, the SM is minimally extended in terms of new particles and couplings, and the gauge group structure is that of the SM. The most minimal incarnation of these simplified models involves two new particles, a particle $X$ playing the role of DM, and a particle $Y$ connecting the DM state to the SM through some new three-point interactions. The spin representations of the new particles provide additional free parameters of the models that generally encompass an additional $\mathbb{Z}_2$ symmetry ensuring DM stability. This is achieved by imposing that all SM states are $\mathbb{Z}_2$-even, and that the DM state is $\mathbb{Z}_2$-odd. In the so-called $s$-channel models~\cite{Fox:2012ru, Haisch:2013ata, Abercrombie:2015wmb, Backovic:2015soa, Boveia:2016mrp}, the mediator is taken $\mathbb{Z}_2$-even whereas in the so-called $t$-channel models considered here it is $\mathbb{Z}_2$-odd. Consequently, $s$-channel mediators couple to both pairs of SM particles and pairs of DM particles, whereas $t$-channel mediators couple to one DM state and one SM state.

In the present work, we consider a class of $t$-channel simplified models for DM in which the mediator couples DM to a right-handed up quark field. Such a class of models features a very simple parameter space with three degrees of freedom once the DM and mediator spin representations are fixed. These parameters consist of the DM mass $m_X$, the mediator mass $m_Y$ and the new physics coupling between them and the up quark. Those models are interesting benchmark scenarios that started to be explored in experimental searches for DM at the LHC~\cite{CMS:2021snz, CMS:2021far}, as well as in the work done through the LHC Dark Matter Working Group.\footnote{See the webpage \url{https://indico.cern.ch/category/16540/}.} This choice is nevertheless only one among all possibilities for $t$-channel simplified models relevant for the LHC, and is motivated by its simplicity and the enhancement of associated collider and direct detection processes due to a connection with valence quarks. Here, the mediator is a state lying in the fundamental representation of $SU(3)_c$ and carrying a hypercharge quantum number of 2/3. We consider the cases in which the mediator $Y$ is a scalar, implying that DM is either a Majorana or a Dirac fermion, and a spin-1/2 fermion, implying that DM is either a scalar or vector state (both of which could either be self-conjugate or not). The spectrum of models covered therefore extends the one investigated in our previous work~\cite{Arina:2020tuw}, which was only dedicated to models featuring self-conjugate DM.

Furthermore, with respect to our previous work, we improve the relic density computation by taking into account Sommerfeld effects relevant in the coannihilation region, and we include additional direct and indirect detection constraints in the analysis of the models. In particular, we apply the latest direct  detection limits from LZ~\cite{LZ:2022ufs}, CRESST-III~\cite{CRESST:2019jnq} and DarkSide-50~\cite{DarkSide:2018bpj}, and reinterpret the indirect detection limits from AMS-02 data on cosmic-ray antiprotons derived in Ref.~\cite{Cuoco:2017iax} within the considered models. Besides, we now cover all scenarios with both real and complex dark matter. In addition, we update LHC constraints by re-interpreting the results of both inclusive and exclusive searches for DM by the ATLAS and CMS collaborations~\cite{CMS:2019zmd, ATLAS:2021kxv, CMS:2021far}. Particular attention is paid to signal modelling. Our work highlights the relevance of same-sign mediator production (see also~\cite{Garny:2013ama}), which has not been considered in any of the experimental analyses including interpretations in $t$-channel DM models, and which turns out to be the driving factor in the determination of LHC constraints in significant parts of the parameter space. Our results therefore point out an important gap in the way the signals have been simulated within $t$-channel models. We provide detailed instructions on how to improve this.

The rest of this work is organised as follows. In section~\ref{sec:model} we briefly introduce the theoretical framework that we use for our study of the six $t$-channel simplified models of DM mentioned above. We refer to \cite{Arina:2020udz} for a more detailed description. Moreover, we additionally provide technical details about the tool chain that is used for both our cosmology and collider investigations. Section~\ref{sec:numerics} is dedicated to our results and the derivation of the most up-to-date bounds on the models considered, first using only cosmological probes (section~\ref{sec:cosmo}) and then only collider probes (section~\ref{sec:colliderpheno}). In section~\ref{sec:cosmoLHCbounds}, we combine these bounds to highlight the strong complementarity between collider physics and cosmology in the exploration of DM models. We conclude and summarise our findings in section~\ref{sec:concl}. This manuscript additionally includes a collection of analytical formulas relevant for DM annihilation in appendix~\ref{app:cross_sections}. 

\section{A unified framework for simplified models of t-channel dark matter}\label{sec:model}

In the present section, we briefly summarise the \dmsimpt\ framework that we use in our study of $t$-channel DM models. Extensive details can be found in~\cite{Arina:2020udz}, and the model files can be obtained online from the \fr\ model database~\cite{feynrulesmodelpage}. Section~\ref{sec:theory} is dedicated to a description of the model itself, and section~\ref{sec:tools} introduces our machinery and how our results have been computed.

\subsection{Theoretical framework}\label{sec:theory}
In any minimal and generic realisation of a $t$-channel simplified model for DM, the field content of the SM is supplemented with a DM candidate $X$, that is taken to be a colourless electroweak singlet. In order to guarantee DM stability, all SM fields are enforced to be even under some {\it ad hoc} $\mathbb{Z}_2$ parity, while the DM state is taken $\mathbb{Z}_2$-odd. In addition, the interactions of the DM with the SM are considered to be mediated by a new state $Y$, that lies in the fundamental representation of $SU(3)_c$ and thus couples to quarks. The mediator $Y$ is imposed to be $\mathbb{Z}_2$-odd, which contrasts with $s$-channel simplified models for DM in which it is $\mathbb{Z}_2$-even~\cite{Fox:2012ru, Haisch:2013ata, Abercrombie:2015wmb, Backovic:2015soa}. In order to maintain generality, we make no assumptions about the spin of the DM and that of the mediator, its representation under the electroweak group, and the flavour structure of its interactions. Consequently, the model gets equipped with a set of 12 mediator fields, one for each flavour and chirality of the SM quarks. Several options for the spins of the $X$ and $Y$ particles are considered. Specific versions of the generic model have been extensively studied in the past (see \eg\ \cite{Garny:2013ama, An:2013xka, Giacchino:2013bta, Bai:2013iqa, DiFranzo:2013vra, Papucci:2014iwa, Garny:2014waa, Garny:2015wea, Ibarra:2015fqa, Berlin:2015njh, Ko:2016zxg, Carpenter:2016thc, Garny:2017rxs, Garny:2018icg, Colucci:2018vxz, Hisano:2018bpz, Mohan:2019zrk, Arcadi:2021glq, Cornell:2021crh, Garny:2021qsr, Becker:2022iso, Belyaev:2022shr, Cornell:2022nky}), but less often in a unified framework as done in~\cite{Chang:2013oia, Giacchino:2014moa, Hisano:2015bma, ElHedri:2017nny, Arina:2020tuw} and in the current work.

The model thus includes three possibilities for the spin of the $X$ particle, that could be a scalar (the complex state $S$ or real state $\tilde S$), a fermion (the Dirac fermion $\chi$ or the Majorana fermion $\tilde\chi$) or a vector (the complex state $V_\mu$ or the real state $\tilde V_\mu$), all those fields being singlet under the SM gauge group $SU(3)_c\times SU(2)_L\times U(1)_Y$. In the case of bosonic DM, the mediator is a fermionic object $\psi$, whereas for fermionic DM, the mediator is a scalar field $\varphi$. The full Lagrangian including the interactions of these fields with the SM can be written as%
\be
 \lag = \lag_{\rm SM} + \lag_{\rm kin} + \lag_{XY}\,,
\ee
where $\lag_{\rm SM}$ is the SM Lagrangian and $\lag_{\rm kin}$ contains gauge-invariant kinetic and mass terms for all new fields. The last term $\lag_{XY}$ includes the interactions of the mediator and the DM with the SM quarks, and could involve a large number of free coupling-strength parameters in the flavour space. 

In order to allow for a tractable phenomenology, we restrict the generic model class to specific cases in which the mediator solely couples to the right-handed up quark, which  we collectively call the {\tt uR} model class. This class is representative of multiple theoretical scenarios. In supersymmetric (SUSY) models, for example, the mixing between squarks which are partners of SM quarks of different chiralities is largely suppressed by the negligible quark masses, and therefore a `right-handed' up squark is allowed to decay to the $u_R$ state and the lightest neutralino. In many SUSY scenarios the latter is a (Majorana fermion) DM candidate~\cite{Ellis:1983ew}, mapping thus the \SM class of models. Furthermore, in models with Universal Extra Dimensions (UED) and conserved Kaluza-Klein (KK) parity, each SM quark is associated with a tower of KK fermionic partners, and the lightest KK-odd of which can decay into SM quarks of definite chirality and the lightest KK-odd state. The latter is usually a DM candidate and can be bosonic ({\it e.g.} a KK-partner of the photon), scalar or vector depending on the UED scenario~\cite{Servant:2002aq}. These models are therefore mapped to the \FS or \FV classes.

Gauge invariance then enforces that the single mediator of the model is an $SU(2)_L$ singlet, and has a hypercharge quantum number of $2/3$. If a left-handed SM quark was chosen, the scenario would have been less minimal, as to ensure gauge invariance with a $SU(2)_L$-singlet DM candidate, a doublet mediator would be needed, implying both up- and down-type components. 

The Lagrangian $\lag_{XY}$ is thus given, in the six setups considered for the DM, by
\be\bsp
   \lag_{XY}^\SM = &\ \lambda\ \overline{\tilde{\chi}} \uR\ \varphi^\dag + {\rm H.c.} \ ,\\
   \lag_{XY}^\SD =&\ \lambda\ \bar \chi \uR\ \varphi^\dag + {\rm H.c.} \ ,\\
   \lag_{XY}^\FS =&\ \lambda\ \bar \psi \uR\  \tilde S + {\rm H.c.} \ ,\\
   \lag_{XY}^\FC =&\ \lambda\ \bar \psi \uR\ S^\dag + {\rm H.c.} \ ,\\
   \lag_{XY}^\FV =&\ \lambda\ \bar \psi \slashed{\tilde{V}} \uR + {\rm H.c.} \ ,\\
   \lag_{XY}^\FW =&\ \lambda\ \bar \psi \slashed{V} \uR + {\rm H.c.} \ .
\esp\ee
Those expressions highlight our notation for the different model possibilities. We denote by \SM (\SD) the model configuration in which the mediator is a scalar state $\varphi$ of mass $M_\varphi$, and the DM is a Majorana (Dirac) fermion $\tilde\chi$ ($\chi$) of mass $M_{\tilde\chi}$ ($M_\chi$). In addition, in \FS (\FC) models the mediator is a fermion $\psi$ of mass $M_\psi$, whereas the DM is a real (complex) scalar state $\tilde S$ ($S$) of mass $M_{\tilde S}$ ($M_S$). Finally, \FV (\FW) models are defined such that the mediator is again a fermion $\psi$ of mass $M_\psi$, but the DM state is this time a real (complex) vector state $\tilde V$ ($V$) of mass $M_{\tilde V}$ ($M_V$). In all these expressions, the coupling of the DM with the mediator and the right-handed up quark is denoted by $\lambda$, regardless of the explicit model configuration.

All the six versions of the model (three with real DM and three with complex DM) depend on three free parameters, namely the DM and mediator masses, and their coupling $\lambda$. In the rest of this work, we will generally denote these three parameters as
\begin{equation}\label{eq:freeparams}
  \big\{ M_X, \ M_Y, \ \lambda \big\} \ ,
\end{equation}
which allows for a unique and model-independent notation in which $M_X$ is the mass of the DM state $X$ and $M_Y$ is the mass of the mediator state $Y$. In addition, we enforce that $M_Y > M_X$ to prevent the DM state to decay into the mediator and an up quark.

The aforementioned theoretical scenarios, and many others, also predict in general interactions with all other SM quark generations, notably with the third one. The phenomenological implications of DM candidates interacting with top or bottom quarks have been extensively studied in the literature, and they are not the subject of the present analysis. Associated predictions are indeed different due to the significantly different quark masses and decay channels, which requires dedicated analysis strategies.


\subsection{Technicalities}\label{sec:tools}

\subsubsection{Tool chain}
The results that are presented in the rest of this work have been obtained with the joint usage of a variety of standard high-energy physics packages, as detailed in \cite{Arina:2020udz}. All associated model files have been obtained from the Lagrangians described in section~\ref{sec:theory}, that have been implemented and processed by \fr~\cite{Christensen:2009jx,Alloul:2013bka}, \nloct~\cite{Degrande:2014vpa} and \fa~\cite{Hahn:2000kx}. This has allowed for the generation of a general next-to-leading order (NLO) UFO~\cite{Degrande:2011ua, Darme:2023jdn} model with five flavours of massless quarks, that we have used within the \mg\ platform~\cite{Alwall:2014hca} for leading order (LO) and NLO computations relevant for the collider phenomenology of the models. Additional LO  model files in which all flavours of quarks are massive have been generated both in the UFO format and in the \ch~\cite{Belyaev:2012qa} format so that they could be used with \micromegas~\cite{Belanger:2013oya,Belanger:2018ccd,Belanger:2020gnr} and \maddm~\cite{Ambrogi:2018jqj,Arina:2020kko,Arina:2021gfn} to assess the cosmology of the six models considered. Non-zero SM quark masses are indeed a necessary ingredient for a reliable calculation of the DM annihilation cross section and direct detection observables.\footnote{In its latest release, \maddm\ has been augmented with the capability of performing automatic tree-induced NLO and loop-induced LO computations from an NLO UFO model~\cite{Arina:2021gfn}. The \dmsimpt NLO UFO models provided on the \fr\ model database can hence be used in \maddm to compute DM annihilation in $\gamma \gamma$ and $gg$ final states. However caution is in order as these UFO models are not compliant for the calculation of electroweak corrections that are potentially relevant for the model's cosmology, and do not feature six flavours of massive quarks.}

\subsubsection{Parameter scan}\label{sec:paramscan}

For each of the six models considered, we perform three-dimensional scans to sample the associated parameter space. We vary the DM and mediator masses and the new physics coupling on logarithmically-spaced grids in the range
\begin{equation}
  M_X, M_Y \in [1, 10^4]~{\rm GeV}\ ,\qquad
  \lambda \in [10^{-4}, 4\pi]\ .
\end{equation}
Furthermore, we require the relative mass splitting $M_Y/M_X-1 \geq 10^{-2}$ as the particular case of a highly compressed mass spectrum is not in the focus of this work.\footnote{For example, in this region of parameter space, collider signatures would originate mostly from the production of long-lived coloured mediators. This would produce bound states, displaced vertices or delayed jets, and thus require dedicated phenomenological and experimental analyses.}  In addition to this sequential grid sampling, we perform a dedicated scan for points matching the measured relic density $\Omega h^2 = 0.12$~\cite{Planck:2018vyg}. To this end, for each configurations of given masses $M_X$ and $M_Y$ we determine the coupling $\lambda$ that allow this criterion to be realised. 

In the next sections, our results are displayed in two-dimensional planes $(M_Y, M_X)$, or alternatively in planes $(M_X, M_Y/M_X-1$). In these cases, the $\lambda$ value is fixed according to three different choices. Either it is calculated so that the amount of DM matches the observed relic density, or it is fixed such that the width-over-mass ratio of the mediator $\Gamma_Y/M_Y$ is equal to a specific value, or it is arbitrarily fixed to a given value. Those choices highlight different aspects of our results. 

\subsubsection{Relic density}\label{sec:relic}
To compute the DM relic density corresponding to a given parameter space point, we assume a scenario in which the DM freezes out. The thermally averaged DM annihilation cross section $\langle\sigma v \rangle$ ($v$ being the relative velocity between the two annihilating particles) is then $d$-wave-suppressed for the real scalar case~\cite{Toma:2013bka, Giacchino:2013bta, Giacchino:2014moa, Giacchino:2015hvk, Biondini:2019int}, $p$-wave-suppressed for Majorana DM~\cite{Giacchino:2014moa, Biondini:2018ovz} and for complex scalar DM, while it proceeds via an $s$-wave for (real and complex) vector DM and for Dirac DM\@. In the case of a velocity-suppressed annihilation cross section $\sigmav$, NLO corrections might be relevant and should therefore be included in the calculation~\cite{Giacchino:2015hvk, Colucci:2018vxz}. In particular, we calculate the loop-induced $X X \to g g$ and $X X\to\gamma \gamma$ annihilation processes, as well as the three-body $X X \to \uR \uRb g$ and $X X \to \uR \uRb \gamma$ channels that could be potentially enhanced by virtual internal bremsstrahlung. In practice, we use the analytic expressions provided in~\cite{Giacchino:2013bta, Giacchino:2014moa, Ibarra:2014qma}, that we have further validated with \maddm. 

The relic density computation can be further refined by including non-perturbative effects such as Sommerfeld enhancement~\cite{Sommerfeld:1931} and bound state formation~\cite{Garny:2021qsr,Becker:2022iso}. While the latter is beyond the scope of this study, we include the former in the computation of the cross section associated with mediator annihilations into gluons ($YY\to gg$) or quarks ($YY\to q \bar q$). This effect is relevant in the co-annihilation regime (see \eg~\cite{Giacchino:2015hvk}).

For a Coulomb potential $V(r)= \alpha/r$ and for an $s$-wave annihilation process (like for mediator annihilations), the Sommerfeld correction factor $S_0$ is defined by~\cite{Cirelli:2007xd,Hannestad:2010zt}
\begin{align}
S_0(\alpha) = -\frac{\pi \alpha/\beta}{1-e^{\pi \alpha /\beta}}\,, \label{eq:Somm}
\end{align}
with $\beta =v /2$.  This expression can be applied to the case of the strong interaction by replacing $\alpha$ with the appropriate factor of $\alphas$. For final states in a pure colour singlet (octet) representation, this gives the replacement $\alpha\to -4\alphas/3$ ($\alphas/6$), the QCD potential being thus attractive (repulsive). As the $gg$ final state can lie either in a colour-singlet or a colour-octet state, the cross section must be decomposed into~\cite{deSimone:2014pda}
\begin{equation}
  S_0(YY\to gg) =  \frac{2}{7}\ S_0(-4\alphas/3) + \frac{5}{7}\ S_0(\alphas/6)\,.
\end{equation} 
In contrast, the $s$-wave annihilation of a pair of mediators in two SM quarks yields
\be 
  S_0 (YY\to q \bar q) = S_0(\alphas/6)\,.
\ee

\subsubsection{Direct detection}\label{sec:DD}
To estimate cross sections relevant for dark matter direct detection, we have made used of \micromegas\ as detailed in \cite{Arina:2020udz,Arina:2020tuw}. This package allows us to evaluate the spin-dependent (SD) DM-nucleon elastic cross section at LO, whereas higher-order QCD correction effects are included for the spin-independent (SI) elastic cross section. In particular, these corrections are crucial for the \SM model as the LO SI cross section vanishes. They are however less significant for other models. Besides the 90\% confidence level (CL) exclusion limits obtained by the null results at the 
LZ experiment~\cite{LZ:2022ufs} for SI elastic scattering and at the PICO-60 experiment~\cite{PICO:2017tgi} for SD elastic scattering, we include in our analysis upper limits on low-mass dark matter stemming from CRESST-III~\cite{CRESST:2019jnq} and DarkSide-50~\cite{DarkSide:2018bpj}. The latter yields in particular competitive exclusion limits on the SI cross section for $M_X$ of 1 GeV or even lower.

\subsubsection{Indirect detection}\label{sec:ID}
Next, we evaluate the limits that apply to the six models considered from observations of gamma-ray lines, gamma-ray continuum, and cosmic-ray antiproton signatures. We impose that the predicted indirect detection signals are compatible with current model-independent exclusion limits at 95\% CL, by combining appropriately the relevant branching ratios into the different annihilation channels.

In the case of the \FS, \SM and \FC models, spectral features in the gamma-ray spectrum are expected to provide the strongest bounds, as tree-level $X X \to \uR \uRb$ annihilations are velocity-suppressed. We therefore derive constraints by considering a combination of $XX$ annihilation into photons and a $\uR\uRb\gamma$ system, the latter being potentially enhanced by virtual internal bremsstrahlung contributions. Using~\maddm~\cite{Arina:2021gfn}, the total annihilation cross section $\sigmav_{\rm tot} = \sigmav_{\uR\uRb\gamma}+2\sigmav_{\gamma\gamma}$ is confronted with the most recent Fermi-LAT~\cite{Fermi-LAT:2015kyq} and HESS~\cite{HESS:2018cbt} data from the Galactic Centre. As bounds obtained by investigating dark matter annihilations into gluons are comparable with bounds arising from gamma-ray line searches, they will not be included in the results presented below. For the \FV, \SD and \FW models, $XX\to\uR\uRb$ annihilations proceed via an $s$-wave configuration. The most stringent indirect detection bounds are thus given by the Fermi-LAT analysis of dwarf spheroidal galaxies (dSph) data~\cite{Fermi-LAT:2016uux} in the $u\bar u$ final state, which we include by using \maddm~\cite{Ambrogi:2018jqj}. Analytic expressions for the annihilation cross section $\langle \sigma v \rangle$ in the various models considered are reported in appendix~\ref{app:cross_sections}.

Additionally, for all six models we study the constraining power of measurements of cosmic-ray antiproton fluxes by the AMS-02 experiment at the International Space Station, as these are expected to provide relevant constraints on dark-matter annihilations in our galaxy. In practice, we employ the results of~\cite{Cuoco:2017iax} and interpret them in the considered models. This analysis derives 95\% CL upper limits on the annihilation cross section as a function of the dark-matter mass for various individual annihilation channels. Moreover, it involves global fits of the cosmic-ray propagation and DM parameters while treating the former as nuisance parameters that are profiled over. 

The models \FV, \SD and \FW all feature a dominant $s$-wave annihilation into a pair of up quarks, so that we can directly apply the limits derived in~\cite{Cuoco:2017iax}. For the \FS, \FC and \SM models, however, $s$-wave annihilation into a pair of first-generation quarks is helicity suppressed and hence virtually absent. Radiation of an extra gluon (or photon) lifts this helicity suppression~\cite{Giacchino:2015hvk,Colucci:2018vxz,Colucci:2018qml}, and loop-induced annihilations into a pair of gluons become relevant. For these model we thus have to consider an admixture of $gg$ and $u\bar u g$ final states, which is not addressed in~\cite{Cuoco:2017iax}. We therefore design a procedure to derive limits.

First, we compute the antiproton source spectra for the channel $XX\to u\bar u g$ with \maddm~\cite{Ambrogi:2018jqj}. The relative contribution from initial bremsstrahlung and final state radiation has a strong dependence on the mass of the mediator, which translates into different angular and momentum distributions at parton level. However, these differences are entirely smeared out after parton showering and hadronisation, such that the antiproton spectrum per annihilation becomes practically insensitive to the mediator mass or the exact spin assignments within a generic $t$-channel DM model. The spectrum is instead, to a good approximation, a function of the sole DM mass. Combining our predictions with the spectrum expected from annihilations into a pair of gluons as provided by the \textsc{PPPC4DMID} package~\cite{Cirelli:2010xx}, we then fit the resulting spectrum with that emerging from all non-leptonic channels for which the analysis in~\cite{Cuoco:2017iax} has been done, after considering the associated DM mass and the normalisation of the combined spectrum as free fit parameters.  Choosing the channel (with its respective best-fit mass $m^\text{ch}_\text{best-fit}$ and normalisation $\zeta^\text{ch}_\text{best-fit}$) that provides the best goodness of fit, we derive limits on the annihilation cross section $\langle\sigma v\rangle_\text{test}^\text{UL}$ associated with the considered (test) spectrum through
\begin{equation}
\langle\sigma v\rangle_\text{test}^\text{UL}
=\langle\sigma v\rangle_\text{ch}^\text{UL}(m^\text{ch}_\text{best-fit}) \, \left(\frac{m_\text{test}}{m^\text{ch}_\text{best-fit}}\right)^2 \frac{1}{\zeta^\text{ch}_\text{best-fit}}\,,
\end{equation}
where $\langle\sigma v\rangle_\text{ch}^\text{UL}(m^\text{ch}_\text{best-fit})$ is the cross-section upper limit for the best-fit channel (evaluated at the best-fit mass).

The above rescaling with the squared mass comes from the fact that the antiproton source term for DM annihilation contains the DM number density squared. Since the analysis in~\cite{Cuoco:2017iax} has been performed for real DM, for the models \FC, \FW and \SD an additional factor $1/2$ needs to be taken into account in the source term, leading to a corresponding weakening of the limit by a factor of 2. The approximate limits derived as described above hold to the extent that the test and best-fit spectra are sufficiently similar. 

To quantify the associated uncertainty, we have repeated the analysis using the channels for which limits have already been derived in~\cite{Cuoco:2017iax}. For each of the channels considered, we have followed the above fitting procedure with the difference of removing the respective channel from the set of reference spectra. Performing this exercise for several channels and masses that yield a similar goodness of fit, we find that the difference between the cross-section limit from our procedure and that of~\cite{Cuoco:2017iax} stays well below 10\%, which provides an estimate of the uncertainty inherent to our procedure.

\subsubsection{Collider bounds}\label{sec:collidersetup}
In order to explore the collider phenomenology of the $t$-channel DM models under study, signal modelling should include three different production channels, namely the production of a pair of DM particles ($p p \to X X$), the production of a pair of mediator particles ($p p \to Y Y + YY^* + Y^*Y^*$), and the associated production of a DM and a mediator particle ($p p \to X Y + XY^*$). Mediator pair production should include contributions originating from QCD-processes (labelled as $YY_{\rm QCD}$ and proportional to $\alphas^2$), $t$-channel DM exchange (labelled as $YY_t$ and proportional to $\lambda^4$) and the corresponding interference (labelled as $YY_i$ and proportional to $\alphas\lambda^2$). Whereas in previous works, the $pp\to YY+Y^*Y^*$ channels, possible when the DM state is a real boson or Majorana fermion, were often ignored (see, however,~\cite{Garny:2013ama,Bai:2013iqa,An:2013xka,Garny:2014waa} for notable exceptions), signal modelling as achieved in this study relies on the full set of diagrams associated with the $t$-channel production of two mediators and anti-mediators. It therefore includes the production of a mediator and an anti-mediator (that interferes with the corresponding QCD diagrams), as well as that of two mediators and two anti-mediators. In the following, the three components of the signal will be generically denoted by $XX$, $YY$ and $XY$. Mediator particles are always enforced to promptly decay into DM particles and quarks, and we always assume a small mediator width so that the narrow-width approximation (NWA) is valid and mediator production and decay factorise~\cite{Berdine:2007uv}. In our exploration of the models' parameter space and in the analysis presented in section~\ref{sec:colliderpheno}, we highlight potential departures from this assumption in specific mass and coupling configurations. In these cases, the results presented may be too crude an approximation.

All our simulations are performed with \mg~\cite{Alwall:2014hca}, using the various \dmsimpt NLO UFO models available online~\cite{feynrulesmodelpage}. The modelling of additional jet emission is essential for the collider phenomenology of the model, its main signature being the production of jets with missing transverse energy carried away by the produced DM particles. It is therefore mandatory to describe it as precisely as possible, {\it i.e.}\ by considering NLO corrections at the matrix-element level. This is achieved for all models, with the exception of those featuring vector DM that rely instead on LO simulations. This originates from the fact that NLO UFO models can only be automatically generated in the Feynman gauge so that we would have issues with the longitudinal degree of freedom of the DM state in the new physics setups with vector DM considered.

We convolve fixed-order NLO matrix elements with the NLO set of NNPDF~3.0 parton distribution functions~\cite{Ball:2014uwa}, driven through the {\sc LHAPDF 6} library~\cite{Buckley:2014ana}. All components of the signal are generated according to the technical details and syntax presented in \cite{Arina:2020udz}. We make use of the {\sc MadSTR} plugin~\cite{Frixione:2019fxg} of \mg\ to appropriately treat the resonant contributions that could emerge from the real corrections to the various processes. We remove the resonant diagram contributions squared for the $YY_t$ and $XY$ channels (corresponding to the {\tt istr=2} option in the {\sc MadSTR} language), and the resonant diagrams themselves for the $XX$ channel to improve the convergence of numerical integration (which corresponds to the {\tt istr=1} option of {\sc MadSTR}). We recall that there is no resonant contribution in the real emission corrections to the $YY_{\rm QCD}$ channel so that there is no need to rely on {\sc MadSTR} in this case. For the production of a pair of mediators ($pp\to YY$) and that of a pair of anti-mediators ($pp\to Y^*Y^*$), the two core files \verb+base_objects.py+ and \verb+loop_diagram_generation.py+ of \mg\ need to be modified in order to allow DM particles to run into virtual diagrams. Moreover, a remaining bug in the method \verb+check_majorana_and_flip_flow+ implemented in the file \verb+helas_object.py+ still needs to be fixed in the most recent publicly available version of \mg (from the 2.9.x series). We refer to the procedure introduced in appendix~A of \cite{Borschensky:2021hbo} for more details. Finally, the modelling of the interference contribution $YY_i$ between the $t$-channel and QCD mediated diagrams relevant to the process $pp\to YY^*$ cannot be achieved in an automated manner at NLO. Consequently, we instead rely on LO simulations and rescale the corresponding cross section with a $K$-factor defined as the geometric mean of the `QCD' and `$t$-channel' $K$-factors,
\begin{equation}
  K_{YY_i} 
    \equiv \sqrt{K_t\ K_{\rm QCD}} = \sqrt{ \frac{\sigma_t^{\rm NLO}}{\sigma_t^{\rm LO}}\
           \frac{\sigma_{\rm QCD}^{\rm NLO}}{\sigma_{\rm QCD}^{\rm LO}} } \ ,
\end{equation}
where $\sigma_x^{\rm LO}$ and $\sigma_x^{\rm NLO}$ denote $pp\to YY^*$ cross sections evaluated at the LO and NLO accuracy in QCD, when restricted to purely QCD diagrams ($x={\rm QCD}$) or purely $t$-channel exchange diagrams ($x=t$). For LO calculations (including any calculation in models with vector DM), the LO set of NNPDF 3.0 parton densities is used.

Mediator decays are handled with {\sc MadSpin}~\cite{Artoisenet:2012st} and {\sc MadWidth}~\cite{Alwall:2014bza} so that off-shell and spin correlation effects are retained. Hard-scattering parton-level events are eventually matched with parton showers, that are modelled by means of {\sc Pythia 8}~\cite{Sjostrand:2014zea} (which additionally handles hadronisation), following the MC@NLO procedure~\cite{Frixione:2002ik}. 

Constraints on the new physics signal emerging from the models considered can be derived from experimental searches for DM in final states comprising jets and missing transverse energy. Such searches can be divided in two categories, namely exclusive searches in which strong constraints are imposed on a small number of jets (see {\it e.g.}\ \cite{ATLAS:2021kxv, CMS:2021far}), and inclusive searches in which loose requirements are enforced instead but on a larger number of jets (see {\it e.g.}\ \cite{ATLAS:2020syg, CMS:2019zmd}). We reinterpret the results of such searches and determine the viable region of the models' parameter spaces, focusing on the analyses of \cite{ATLAS:2021kxv, CMS:2019zmd, CMS:2021far} for which implementations in public tools exist~\cite{Araz:2020lnp, Mrowietz:2020ztq, Fuks:2021zbm, CMS:2021far}.\footnote{Codes available from \url{https://doi.org/10.14428/DVN/4DEJQM}, \url{https://doi.org/10.14428/DVN/IRF7ZL} and \url{https://doi.org/10.14428/DVN/4TGJAV}.} In practice, we make use of the \ma\ framework~\cite{Conte:2012fm, Conte:2014zja, Conte:2018vmg}, which relies on {\sc FastJet} (version 3.3.4)~\cite{Cacciari:2011ma} and its implementation of the anti-$k_T$ algorithm~\cite{Cacciari:2008gp}, {\sc Delphes~3}~\cite{deFavereau:2013fsa}, and the SFS framework~\cite{Araz:2020lnp} for the simulation of the LHC detectors. The more stringent constraints being those obtained through the recast of the CMS-SUS-19-006~\cite{CMS:2019zmd} search, only the corresponding predictions will be displayed in the next section, together with naive extrapolations at 300~fb$^{-1}$ and 3~ab$^{-1}$ relevant for the current run (run~3) and the high-luminosity (HL-LHC) run of the LHC in~section~\ref{sec:cosmoLHCbounds}. The latter have been derived following the methodology of \cite{Araz:2019otb}. As such an inclusive analysis was omitted from our previous study~\cite{Arina:2020tuw}, this work updates all previous constraints, leading to much stronger restrictions on the viable part of the models' parameter spaces. Results are further detailed in the next section.

In addition, measurements of the $Z$-boson visible decay width provides robust bounds on the models for scenarios in which the mediator is lighter than half of the $Z$-boson mass, as in this case it can be pair-produced through a $Z$ decay. We refer to \cite{Zyla:2020zbs} for extra details.

\section{Numerical results}\label{sec:numerics} 

\subsection{Cosmological and astrophysical constraints}\label{sec:cosmo}

Cosmological and astrophysical constraints provide important guidance on the potential relevance of the regions of the parameter spaces of the models that we probe. We present implications of those constraints in two classes of $t$-channel DM scenarios in which either the $X$ state makes up all of the observed DM (i), or when it only accounts for a fraction of it (ii). The first condition maps out a two-dimensional hypersurface of the three-dimensional parameter space defined by the parameters shown in~\eqref{eq:freeparams} (\emph{cf.}~section~\ref{sec:paramscan}), whereas the latter only provides, for a given mass configuration, a lower bound on the coupling arising from an over-closure constraint. While both choices lead to relevant benchmark scenarios, the cosmological constraints derived could nevertheless generally be softened further by considering, for instance, a non-standard cosmological history and/or an extended particle content.

\begin{figure*}
\centering
\includegraphics[width=.45\textwidth]{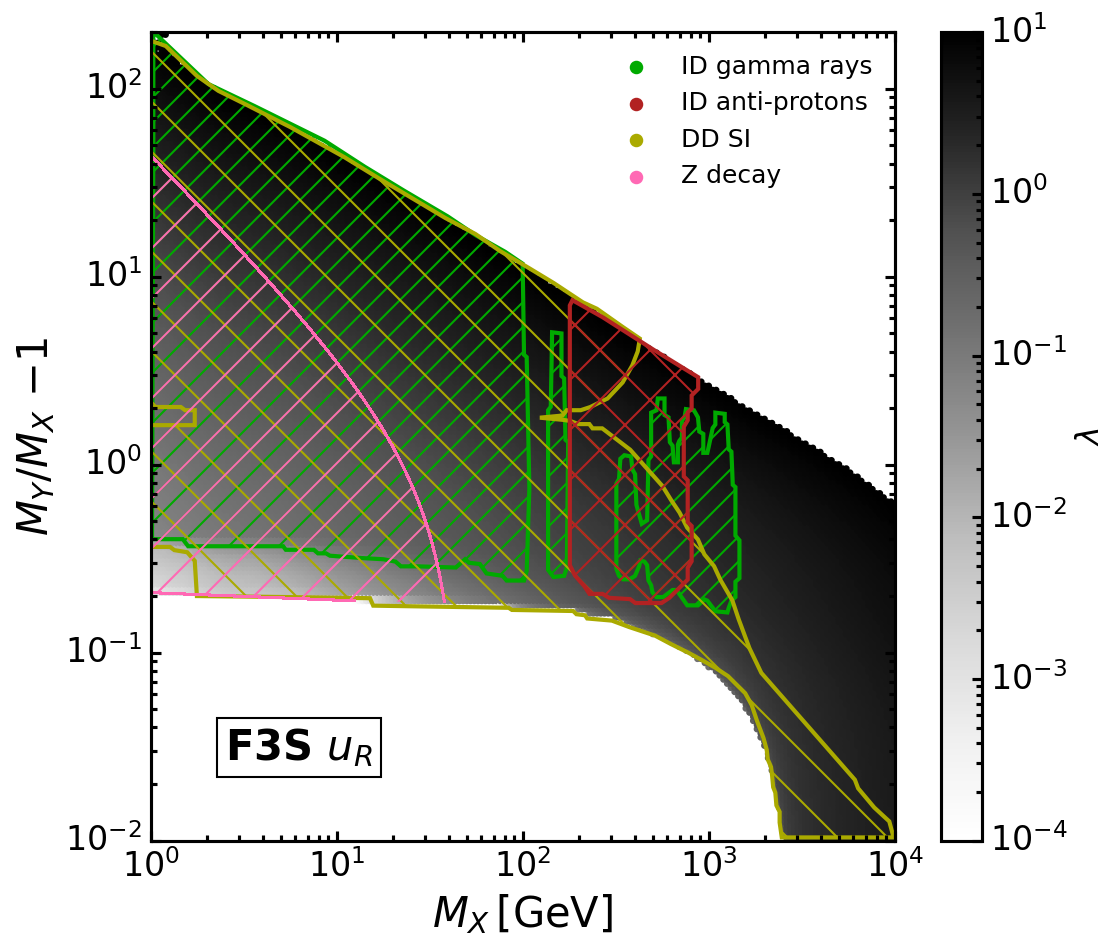}\hspace{3ex}
\includegraphics[width=.45\textwidth]{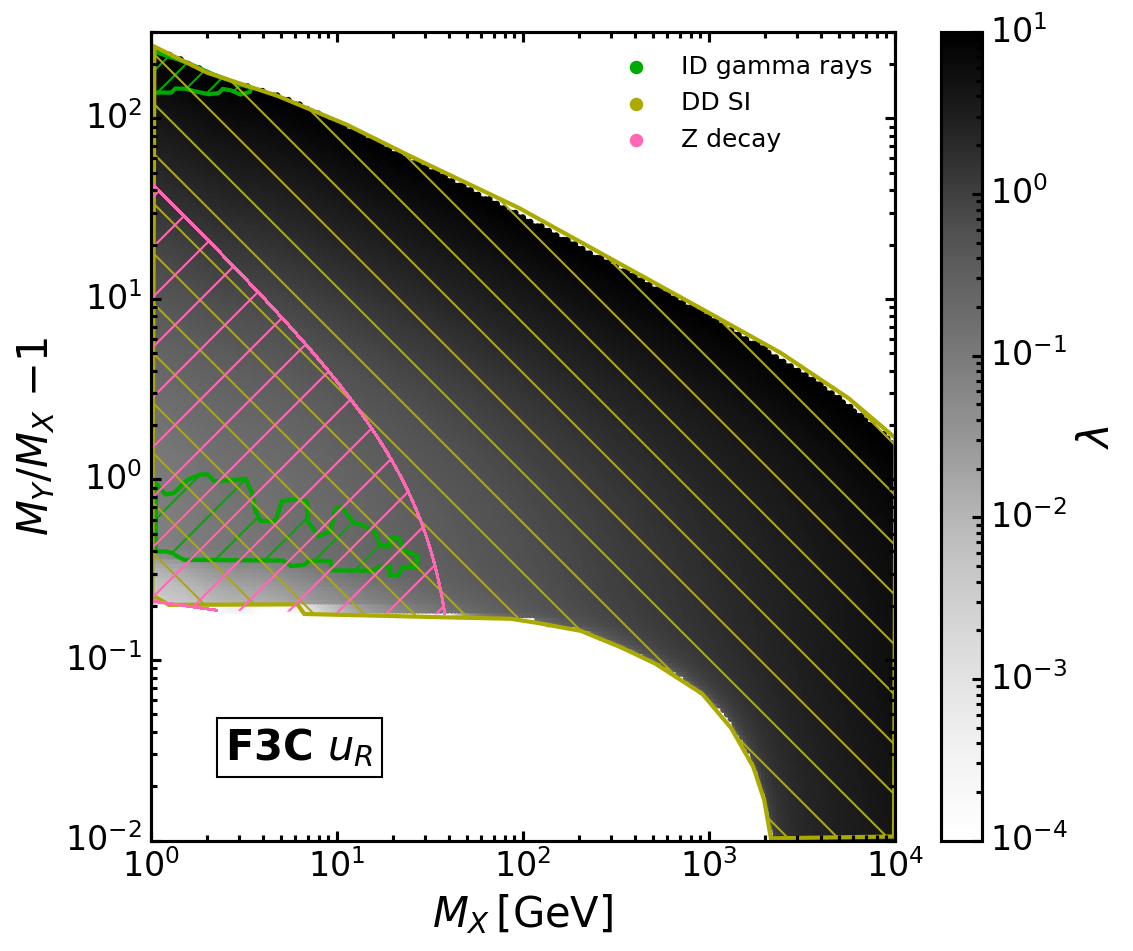}\vspace{1ex}

\includegraphics[width=.45\textwidth]{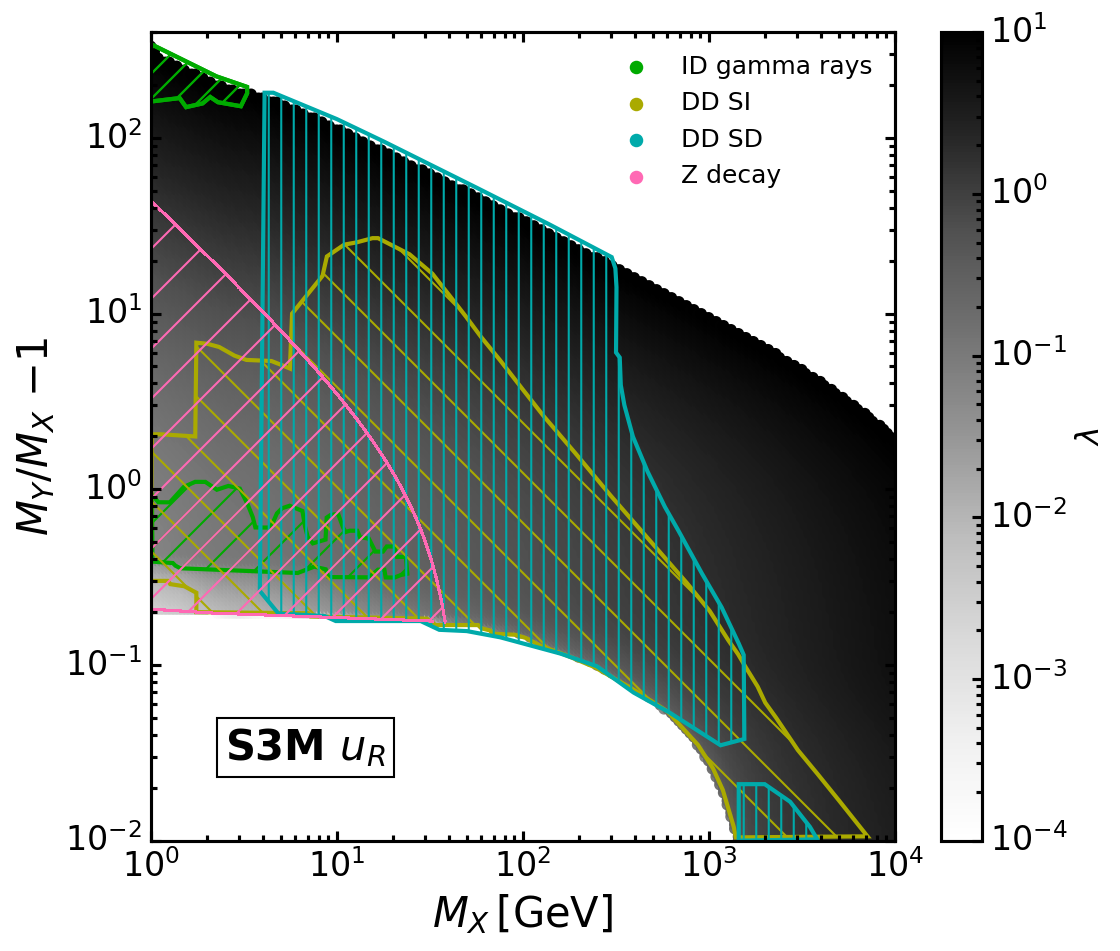}\hspace{3ex}
\includegraphics[width=.45\textwidth]{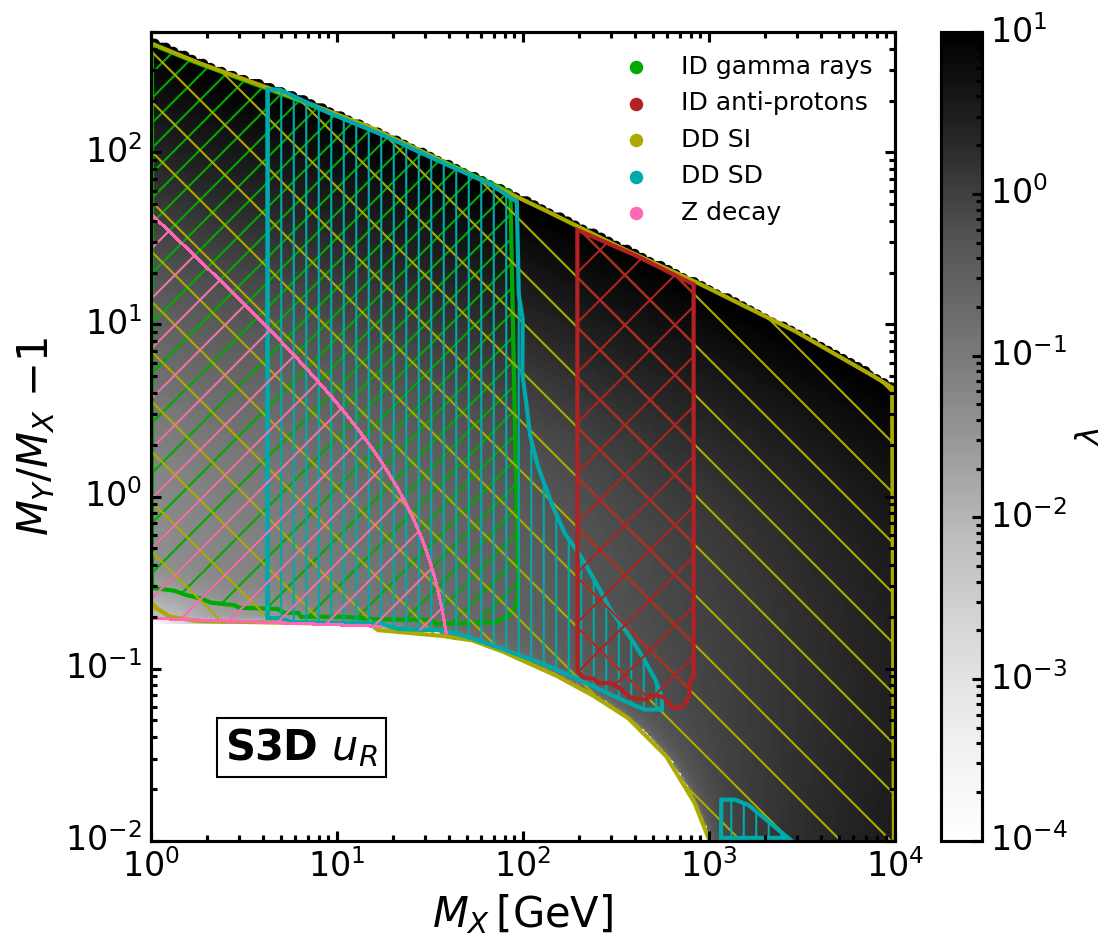}\vspace{1ex}

\includegraphics[width=.45\textwidth]{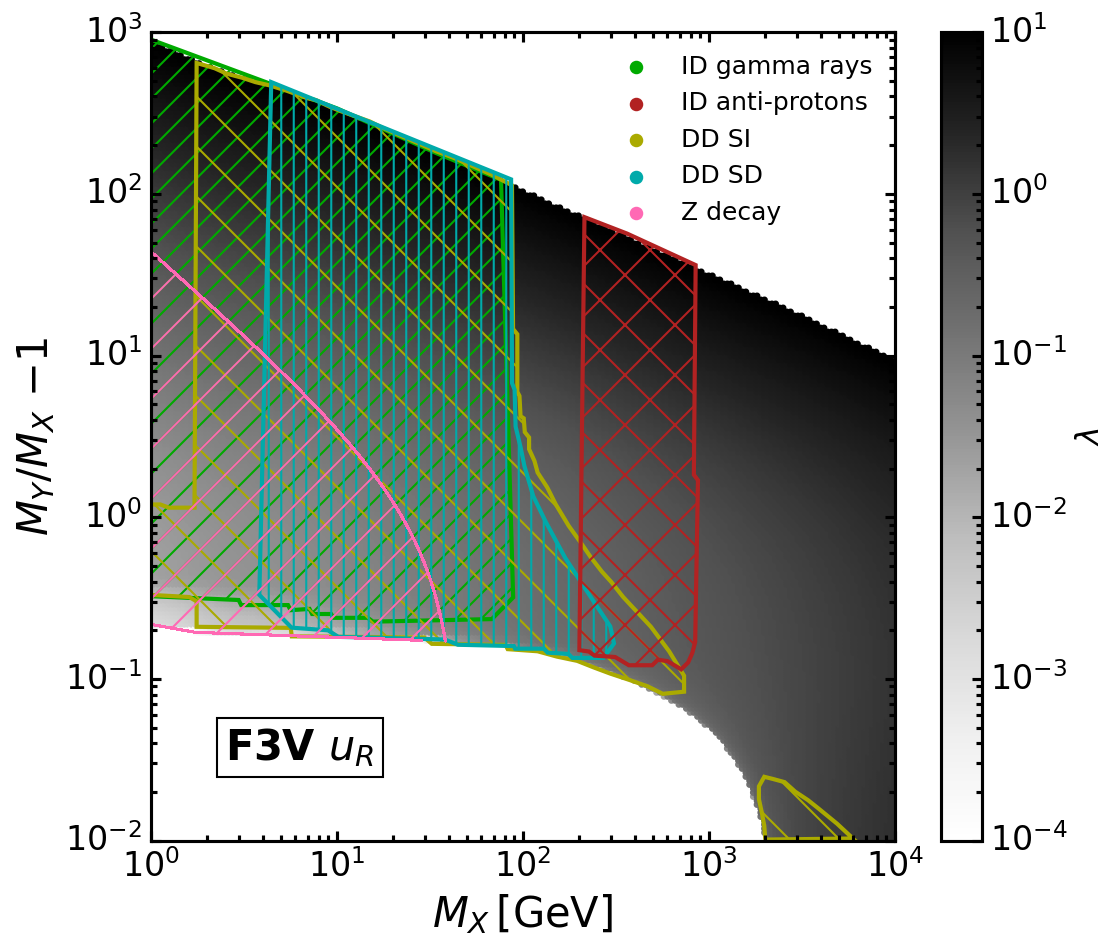}\hspace{3ex}
\includegraphics[width=.45\textwidth]{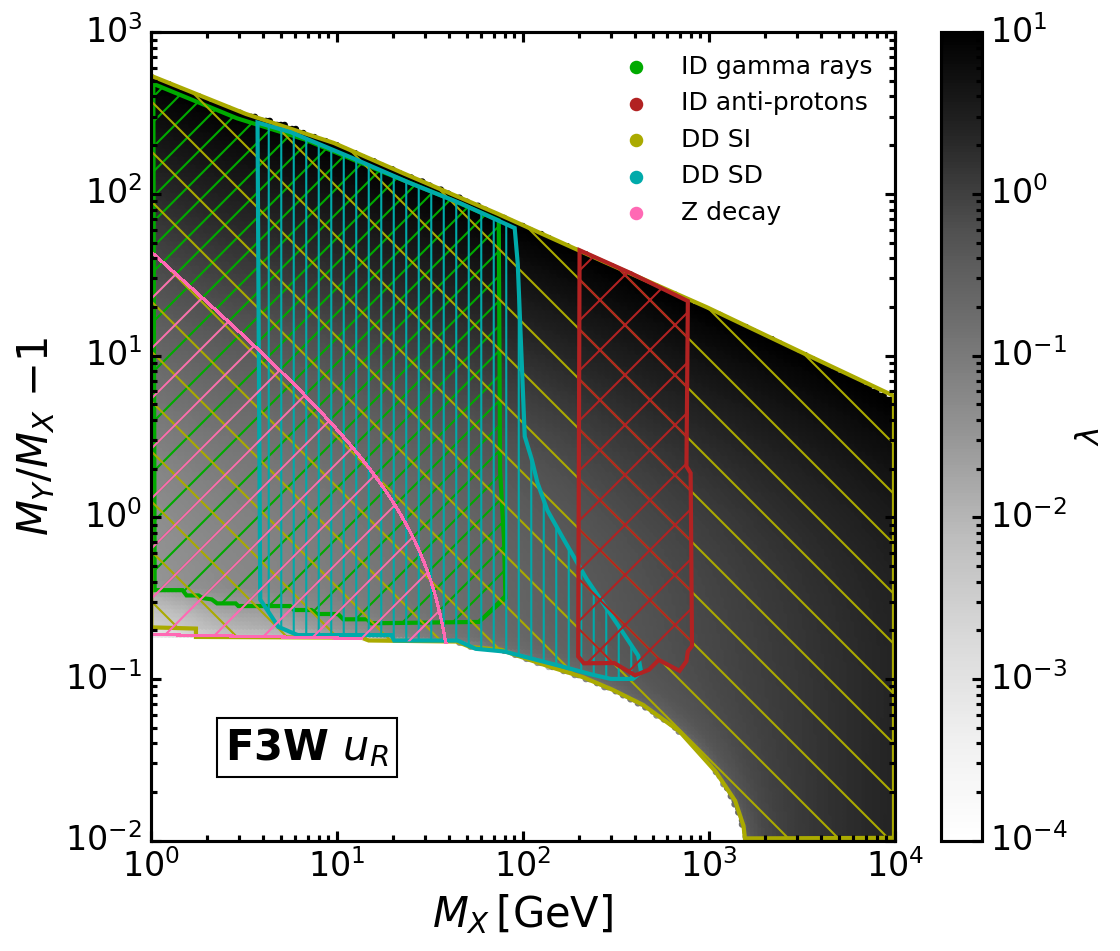}
\caption{\label{fig:cosmo3D}Constraints on the $t$-channel DM models investigated that emerge from cosmological and astrophysical observables, as well as from the measured $Z$-boson visible decay width. The coloured hypersurfaces displayed in the different $(M_X, M_Y/M_X-1)$ planes correspond to scenarios that satisfy $\Omega h^2\simeq0.12$ for a value of the coupling $\lambda$ reflected by the grey-scale colour map. The left (right) panels correspond to models with self-conjugate (complex) DM, and we consider a scalar (top row), fermion (central row), and vector (bottom row) DM candidate. The hatched regions denote exclusions from gamma-ray searches (`ID gamma rays'), searches in cosmic-ray antiprotons (`ID anti-protons'), DM direct detection via spin-independent and spin-dependent interactions (`DD SI' and `DD SD', respectively), and $Z$-boson visible decays (`Z decay'). For details we refer to sections~\ref{sec:ID}, \ref{sec:DD} and to the end of section~\ref{sec:collidersetup}.}
\end{figure*}

In figure~\ref{fig:cosmo3D} we show the cosmologically viable parts of the parameter space in the case (i) for all six models considered. The grey-scale colour maps reflect the value of the coupling $\lambda$ that allows for an explanation of the measured DM relic density $\Omega h^2\simeq0.12$~\cite{Planck:2018vyg}, and our results are displayed in the plane spanned by the dark matter mass $M_X$ and the relative mass splitting, $M_Y/M_X-1$. 

Toward large DM masses and mass splittings, the coupling value required to predict a DM relic density in agreement with Planck data increases. The white area visible in the upper right corner of the different panels consequently does not correspond to scenarios leading to a sufficiently large annihilation cross section within the perturbative regime of the coupling. We notice that for these observables the mediator width does not play a role, as the mediator always propagates non-resonantly either due to the $t$-channel topologies or to the energy scales involved. Toward small mass splittings, coannihilation effects of the mediator particle become increasingly important. In the white area shown in the lower left corner of the panels, mediator pair annihilation yields an annihilation cross section that is so large that it alone leads to under-abundant DM provided that the DM and mediator states are in chemical equilibrium. While this condition is met for the considered values of the coupling $\lambda\ge 10^{-4}$, cosmological viable solutions can also be found for couplings of the order of $10^{-6}$. In this case, the relic density is set by conversion-driven freeze-out~\cite{Garny:2017rxs}, in which the above chemical equilibrium breaks down due to semi-efficient conversion processes between the DM particle and the mediator. The computation of the precise coupling value matching $\Omega h^2\simeq 0.12$ in such a scenario requires to solve a coupled set of Boltzmann equations, which is beyond the scope of this work. We nevertheless emphasise that as $\lambda<10^{-4}$, all astrophysical and collider constraints discussed below are evaded. 

Direct detection bounds originating from SI DM interactions with nuclei are among the strongest bounds that could be imposed on all six models. For the considered case (i), the interplay of predicting a relic density in agreement with data and considering scenarios viable relatively to direct detection constraints already excludes the entire sampled parameter space for all three complex DM scenarios, as shown in the right panels of figure~\ref{fig:cosmo3D}. This leaves the conversion-driven freeze-out region as the only allowed region within those models (within the framework of frozen-out DM considered here). For self-conjugate DM models, parts of the parameter space are not challenged by direct detection constraints. However, the combination of direct detection bounds via SI and SD interactions, indirect detection via gamma-ray and cosmic-ray antiproton probes, as well as the robust constraints emerging from $Z$-boson visible decay measurements, excludes large parts of the considered parameter spaces, as depicted in the left panels of figure~\ref{fig:cosmo3D}. For real scalar DM, it hence excludes the entire region with DM masses below 800\,GeV or mediator masses below  2\,TeV. For Majorana fermion and real vector DM the situation is similar, with the exception of additional allowed islands of scenarios when $M_X\lesssim 4$ GeV and $100\,\text{GeV}\lesssim M_X \lesssim 200\,$GeV respectively. For the latter class of scenarios, we can note that cosmic-ray antiproton data provide important limits in the range $200\,\text{GeV}\lesssim M_X \lesssim 800\,$GeV, as for scalar DM. This is due to the fact that indirect detection bounds are particularly strong for (real and complex) vector and Dirac fermion DM, since the annihilation into quark pairs is mediated by an $s$-wave process.

We now move on to the second class of scenarios considered, namely case (ii), in which the DM is made of several components. The measured relic density therefore only consists of an upper limit on the theoretical predictions made in the context of the $t$-channel DM models studied. As a consequence, additional regions of the parameter space open up. In order to assess the constraints that could be imposed on such scenarios, we assume that $X$ only accounts for a fraction of the DM, and we subsequently rescale the predicted direct (indirect) detection signals by that fraction (squared). By doing so, we implicitly assume that the local and global DM composition is equal, \emph{i.e.}~that the clustering properties of the different DM components do not differ significantly. In the following, we show the corresponding results in the $(M_Y, M_X)$ plane, for either a fixed value of the mediator width-over-mass ratio $\Gamma_Y/M_Y$, or for a fixed $\lambda$ value. Such a way to present our results will allow for a direct comparison with the collider constraints derived in section~\ref{sec:colliderpheno}.  For the same reason, we furthermore restrict our analysis to the case of self-conjugate DM, which is the only option to get viable ({\it i.e.}\ non excluded) regions of the parameter space that are testable at the LHC under the assumption of prompt mediator decays.

\begin{figure*}
\includegraphics[width=.325\textwidth]{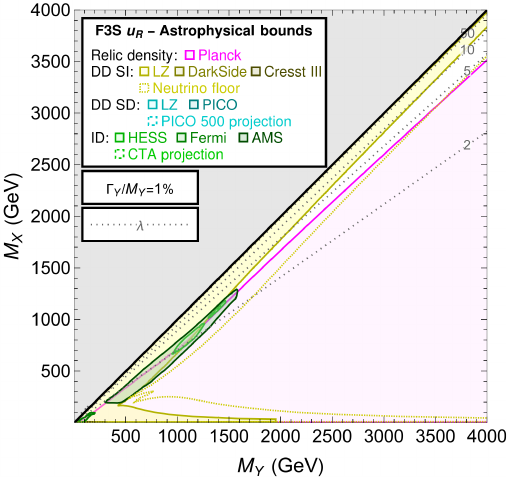}
\includegraphics[width=.325\textwidth]{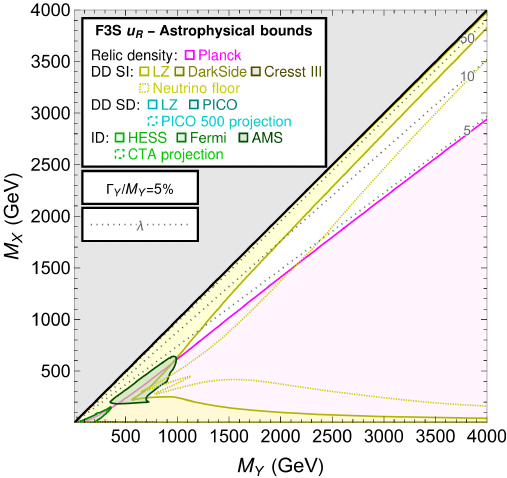}
\includegraphics[width=.325\textwidth]{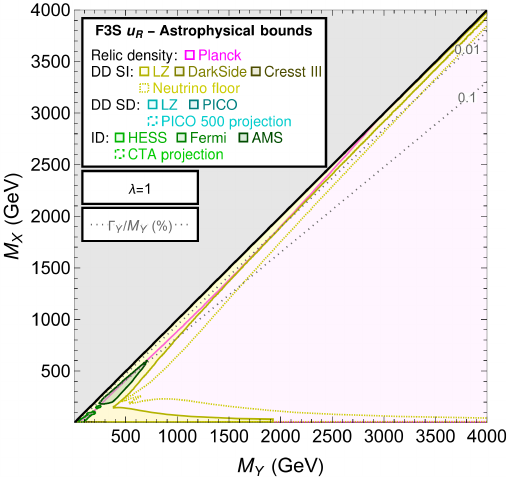}
\includegraphics[width=.325\textwidth]{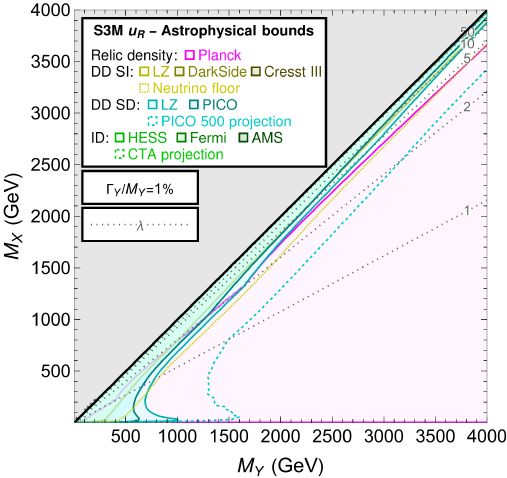}
\includegraphics[width=.325\textwidth]{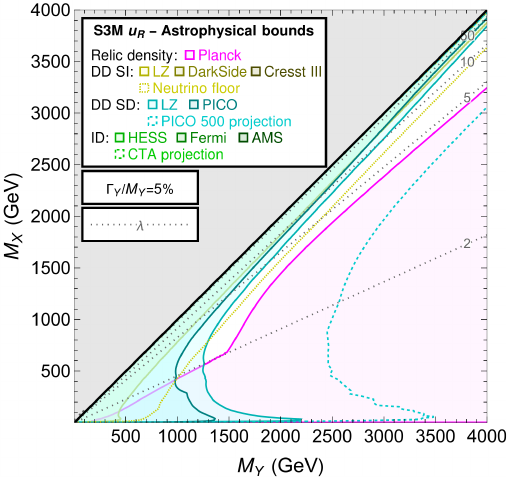}
\includegraphics[width=.325\textwidth]{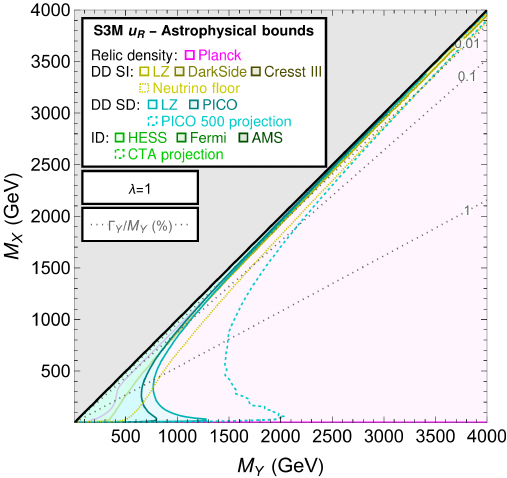}
\includegraphics[width=.325\textwidth]{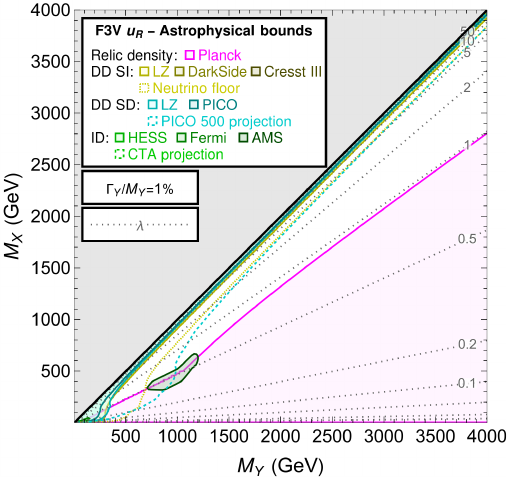}
\includegraphics[width=.325\textwidth]{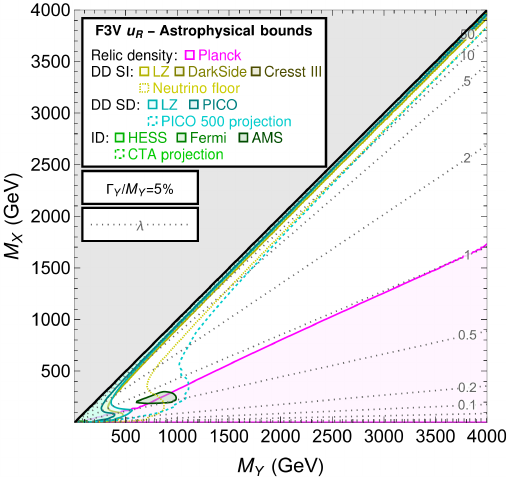}
\includegraphics[width=.325\textwidth]{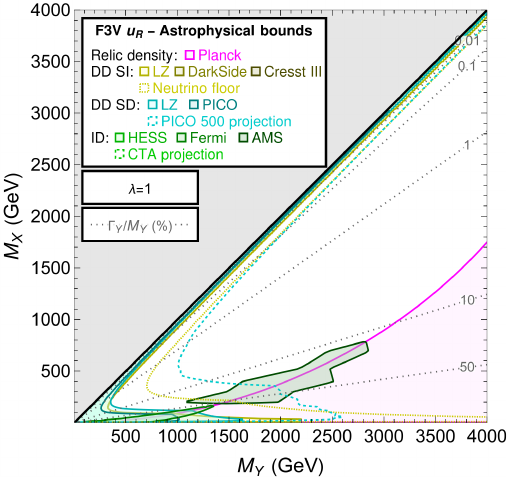}
\caption{\label{fig:3DcosmoWoM}Cosmological and astrophysical bounds on the real DM $t$-channel model considered, as obtained from different observables for the \FS (top row), \SM (middle row) and \FV (bottom row) class of scenarios. The results are shown in the $(M_Y,M_X)$ plane for $\Gamma_Y/M_Y=0.01$ (left column) and 0.05 (middle column), as well as for $\lambda=1$ (right column). Shaded areas are excluded by SI (yellow) and SD (teal) direct detection probes, as well as by indirect (green) detection searches. Scenarios featuring a relic density $\Omega h^2=0.120$ are reflected by the border of the exclusion originating from Planck data (shown in purple). Finally, grey dotted lines are isolines of constant $\lambda$ values (left and middle columns) or of constant $\Gamma_Y/M_Y$ value (right column).}
\end{figure*}

In the left and middle panels of figure~\ref{fig:3DcosmoWoM}, we show the interplay of cosmological and astrophysical constraints on scenarios in which the mediator width-to-mass ratio is fixed to a specific value. We adopt $\Gamma_Y/M_Y=0.01$ (left column) and $0.05$ (central column). The purple area shown in the nine subfigures consists of the regions of the \FS (top row), \SM (central row) and \FV (bottom row) parameter spaces in which DM is over-abundant. Scenarios in which the relic density match the measured value therefore lie at the boundary of the Planck exclusion regions, and they are thus represented by dark purple lines. Conversely, parameter space regions displayed through the various white areas correspond to regions in which DM is under-abundant, and that are additionally allowed by all astrophysical constraints considered. The associated exclusion are shown through yellow, teal and green exclusion contours for SI direct detection constraints, SD direct detection constraints, and indirect detection constraints, respectively. Generally, it turns out that for all models the under-abundant regions become larger with increasing values of $\Gamma_Y/M_Y$, due to the larger couplings involved.

The importance of the individual astrophysical bounds varies strongly with the spin of the mediator and that of the DM state, as already discussed in the context of figure~\ref{fig:cosmo3D}. In addition to current constraints, we further display in figure~\ref{fig:3DcosmoWoM} the projected direct and indirect detection sensitivity of the PICO-500~\cite{PICO:2016kso,Kang:2018odb} and CTA~\cite{Lefranc:2016fgn,CTA:2020qlo} experiments, respectively, as well as the sensitivity corresponding to the so-called `neutrino floor'~\cite{Billard:2013qya} limiting the direct detection reach in the foreseeable future. This allows for an assessment of the improvement, in terms of coverage of the parameter space of the models, that could be expected from future astrophysical probes. Whereas such an improvement is mild for models with $\Gamma_Y/M_Y=0.01$, the larger $\lambda$ coupling values inherent to scenarios with $\Gamma_Y/M_Y=0.05$ makes it more significant.

In the right column of figure~\ref{fig:3DcosmoWoM}, we show the corresponding results for scenarios in which the new physics coupling is fixed to the specific value $\lambda=1$. Both in the real scalar and Majorana DM case, the entire set of mass configurations probed is excluded by relic density constraints. In such setups in which $\lambda$ is fixed, the parameter space only opens up for very large coupling values like $\lambda\gtrsim 4.8$ and 3.5 in the \FS and \SM cases respectively, the size of the excluded areas shrinking drastically with increasing $\lambda$ values larger than these thresholds.

\subsection{Collider phenomenology}\label{sec:colliderpheno}
In this section, we follow the simulation strategy described in section~\ref{sec:collidersetup} and present predictions for two classes of scenarios defined in~\ref{sec:paramscan}. Both of these consist of scenarios in which the two new physics masses, $M_X$ and $M_Y$, are free. In the first set of scenarios, the coupling $\lambda$ is chosen such that the mediator width-over-mass ratio $\Gamma_Y/M_Y$ is fixed to a given value (section~\ref{sec:collider_GoverM}). For the second class of scenarios, the coupling $\lambda$ is instead fixed to an arbitrary value (section~\ref{sec:collider_lam}).

\subsubsection{Scenarios with fixed $\Gamma_Y/M_Y$ ratio}\label{sec:collider_GoverM}

\begin{figure*}[t]
\includegraphics[width=.32\textwidth]{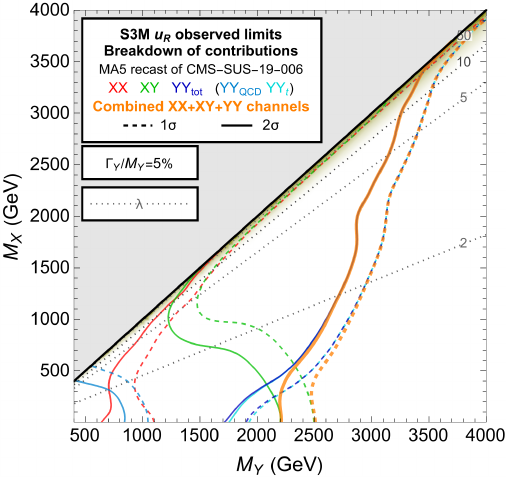}
\includegraphics[width=.32\textwidth]{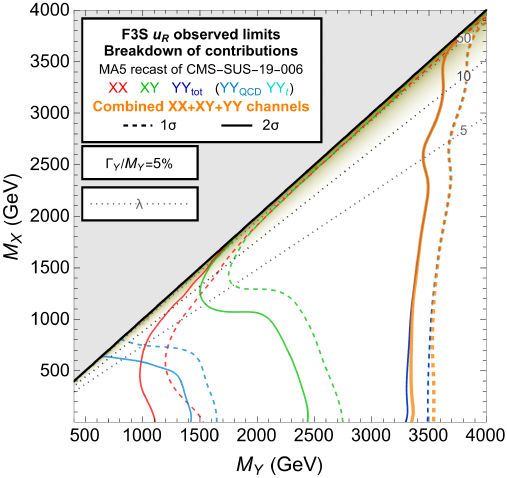}
\includegraphics[width=.32\textwidth]{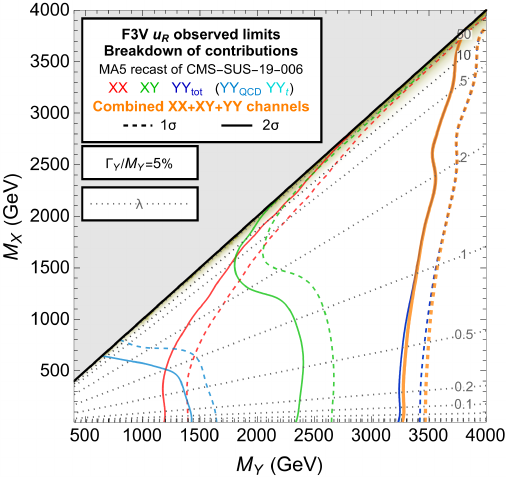}
\caption{\label{fig:ClsContributions} Exclusion limits at 68\% confidence level (1$\sigma$, dashed lines) and 95\% confidence level (2$\sigma$, solid lines) from the reinterpretation of the results of the CMS search~\cite{CMS:2019zmd}. We consider the three real DM scenarios described in section~\ref{sec:theory}, namely \SM (left), \FS (centre) and \FV (right), and a fixed ratio $\Gamma_Y/M_Y=0.05$. The dotted grey lines are isolines of constant $\lambda$ coupling value, and the excluded regions of the parameter space lie at the left of the orange lines. The bounds from individual processes $XX$ (red), $XY$ (green) and $YY$ (dark blue) are also shown to illustrate their relative role, the $YY$ process being further split into its purely QCD part ($YY_{\rm QCD}$, teal), and its purely $t$-channel part ($YY_t$, turquoise). The area highlighted through the yellow gradient indicates when the coupling becomes so large that a perturbative approach becomes less and less valid.}
\end{figure*}

As written above, for each mass configuration $(M_Y, M_X)$ considered we derive the value of the $\lambda$ parameter that leads to a given mediator width-over-mass ratio $\Gamma_Y/M_Y$. We then simulate events with the machinery of~section~\ref{sec:collidersetup}, and next recast the results of the CMS-SUS-19-006 analysis~\cite{CMS:2019zmd} of the full LHC run~2 dataset to determine bounds on the model. The exclusion limits that we obtain for $\Gamma_Y/M_Y = 5\%$ are shown in figure~\ref{fig:ClsContributions} for the three real DM scenarios considered, namely the \SM (left panel), \FS (central panel) and \FV (right panel) models. As the coupling between the mediator, the DM, and the up-type quark is a function of the two new masses $M_X$ and $M_Y$, we additionally display in all figures isolines of constant $\lambda$ values (grey dotted lines). Scenarios for which the obtained value of $\lambda$ is larger than 10 are highlighted through a yellow gradient. For such model configurations, any prediction should however be interpreted very carefully. The whole collider approach adopted in this work indeed relies on a perturbative treatment of the amplitudes of the different involved processes that is only valid for moderate coupling values $\lambda$ well below $4\pi$. Moreover, we recall that scenarios with non-self-conjugate DM are excluded by cosmological constraints, as detailed in~section~\ref{sec:cosmo}, and are thus not relevant in light of searches at colliders. They are therefore ignored in the present discussion.

In order to assess whether a given benchmark is phenomenologically allowed, we rely on the CL$_{\rm s}$ method~\cite{Read:2002hq}, and make use of the number of events expected from the SM background (as publicly provided by the CMS collaboration), the number of observed events (also provided by the CMS collaboration), as well as of the number of signal events that we predict. Our recast considers all 174 signal regions of the CMS-SUS-19-006 analysis, together with the 12 aggregate search regions targeting specific signal topologies. However, limit extraction does not only conservatively rely on the most sensitive of all search regions, but also exploits the fact that the CMS public results include correlation information in the form of an approximate covariance matrix. Signal regions can consequently be combined under the assumptions that systematic uncertainties in signal modelling can be neglected, and that uncertainties on the background contributions are Gaussian~\cite{CMS-NOTE-2017-001}. The corresponding combination procedure is available in an automated fashion within \ma for about a year~\cite{Alguero:2022gwm}.

The exclusion bounds at 95\% confidence level derived when the `{\it full}' new physics signal (including the channels $XX$, $YY$ and $XY$) is accounted for, namely the processes
\begin{equation}
    p p \to XX\,, \quad XY + XY^*\,,\quad YY + YY^* + Y^*Y^*\ ,
\end{equation}
are shown through solid orange lines, the corresponding exclusions at $68\%$ confidence level being represented by dashed orange lines. Scenarios lying on the left of the lines are excluded. The most striking feature of the exclusion bounds presented in figure~\ref{fig:ClsContributions} is that they are much higher than those found in our previous study~\cite{Arina:2020tuw}. In the latter earlier work, we reinterpreted the results of an analogous inclusive multi-jet plus missing transverse energy search \cite{ATLAS:2019vcq} (whose results have been in the meantime peer-reviewed in \cite{ATLAS:2020syg}), and obtained bounds of 1.5--2~TeV in the \SM model and of 2--2.5~TeV in the \FS and \FV models, regardless of the DM mass. In figure~\ref{fig:ClsContributions}, those bounds are more stringent. They reach 2.2--3.7~TeV in the \SM model and 3.3--3.8~TeV in the \FS and \FV models, as depicted by the solid orange lines in the left, central and right panel of the figure. 

In order to understand the origin of this improvement, we break down the signal into its different contributions, also shown in figure~\ref{fig:ClsContributions}. The bounds given by red lines are those determined when only the $pp\to XX$ channel contributes to the signal, whereas those represented by green lines refer to a signal only emerging from the $pp\to XY + XY^*$ process. Limits obtained by solely considering \mbox{(anti-)mediator} pair production are given by the various blue lines. The darkest shade of blue corresponds to a signal including all (QCD and $t$-channel exchange) diagrams ($YY_{\rm tot}$), whereas the teal lines are dedicated to a signal in which only QCD diagrams are included ($YY_{\rm QCD}$). Finally, the turquoise lines (always almost completely coinciding with the $YY_{\rm tot}$ lines) refer to a signal only including $t$-channel DM exchanges. For all individual processes solid lines are again used for exclusions at 95\% confidence level, and dashed lines refer instead to exclusions at 68\% confidence level. We observe that the $YY_t$ bounds are superimposed with the `full' bounds in most parts of the parameter space, the only exception being for the \SM model in the low-mass DM region where the channel $XY$ becomes dominant. This $YY_t$ dominance is due to the contribution of the $uu$-initiated partonic process, $uu\to Y Y$, that proceeds via $t$-channel DM exchange and whose cross section is enhanced by the potential presence of two valence quarks in the initial state (see {\it e.g.}~\cite{Garny:2013ama}). This contribution is peculiar to real DM scenarios as in complex DM scenarios, it is impossible to produce a pair of mediators or of anti-mediators (the only contributing process being $u\bar u \to Y Y^*$). However, such a $uu$-initiated process has not been included in the experimental analyses of $t$-channel DM models (as well as in our previous study~\cite{Arina:2020tuw}). Remarkably, its relevance is way larger than that of the QCD contribution to the $YY$ channel, that is also initiated by $u\bar u$. Focusing on a signal driven by QCD-induced mediator-anti-mediator pair production ($pp\to YY^*$), mediators are enforced to be heavier than about 800~GeV in the \SM model and 1.5~TeV in the \FS and $\FV$ models, the bounds first decreasing with DM masses increasing up to 500-700~GeV before vanishing entirely.
It is worth noting that an analogous enhancement would be achieved by considering scenarios where the DM interacts with down quarks. However, due to the smaller contribution of down parton densities, the numerical relevance of same-sign mediator production would be smaller than in the up-quark case.

\begin{figure*}
\includegraphics[width=.32\textwidth]{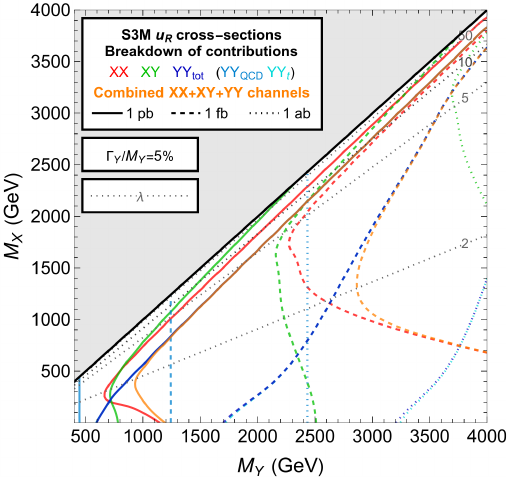}\hfill
\includegraphics[width=.32\textwidth]{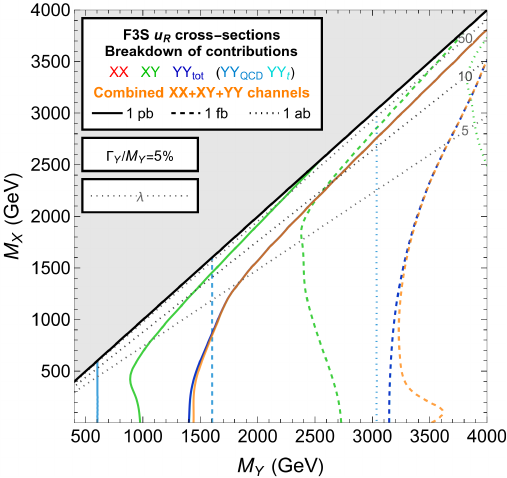}\hfill
\includegraphics[width=.32\textwidth]{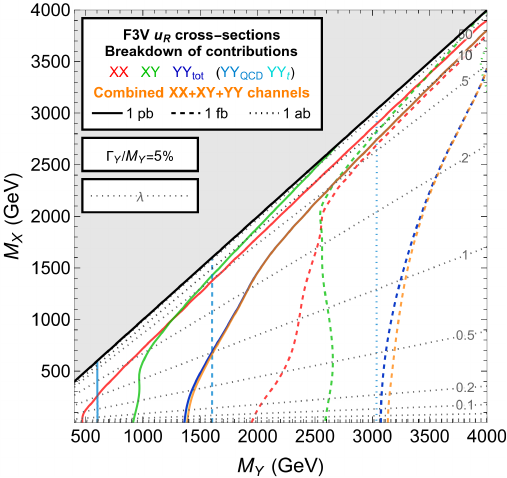}
\caption{\label{fig:XS} Contours of constant cross section for the three real DM scenarios introduced in~section~\ref{sec:theory}, namely \SM (left), \FS (centre) and \FV (right), and a fixed ratio $\Gamma_Y/M_Y=0.05$. The dotted grey lines are isolines of constant $\lambda$ coupling value, and the orange lines show the total cross section combining the contributions of the three individual processes $XX$, $XY$ and $YY$. The corresponding individual cross sections are given through red, green and dark blue lines respectively, the $YY$ contribution being split according to its pure QCD (teal) and $t$-channel component (turquoise).}
\end{figure*}

A further noticeable feature of the bounds obtained in the three models is that they (relatively) weakly depend on $M_X$, especially in scenarios with a fermionic mediator. This is rather counter-intuitive, as in a class of scenarios with a fixed $\Gamma_Y/M_Y$ value, for constant $M_Y$ the coupling should decrease with $M_X$. This behaviour is illustrated by the grey isolines in figure~\ref{fig:ClsContributions} that represent sets of scenarios sharing a common $\lambda$ coupling value. To understand this point, it is instructive to factorise out any effect stemming from the recasting procedure (experimental efficiencies, detector effects, {\it etc.}) and investigate analytically the partonic cross sections associated with each of the processes considered. In figure~\ref{fig:XS} contours of constant cross sections are plotted in the $(M_Y, M_X)$ plane. For fermionic mediator scenarios (\FS and \FV in the central and right panel of the figure) the cross section is always dominated by the $uu$-initiated $YY_t$ process. In contrast, in the scalar mediator scenario (\SM, left panel of the figure) the $XX$ channel contributes more significantly and is even dominant for small DM masses. Nevertheless, the corresponding experimental efficiencies make it negligible in the determination of bounds. The selection in the CMS search that we recast~\cite{CMS:2019zmd} indeed enforces the presence of multiple hard jets in the final state, the signal regions driving the exclusion generally requiring two or three jets, a large hadronic activity $H_T$ and a large amount of missing transverse energy. Consequently, the $XX$ channel, that mostly leads to the production of a small number of hard jets, is not so relevant in terms of potential constraints on the model.

We therefore focus on the $YY_t$ channel only in the rest of this section. Both exclusion levels and associated cross sections (blue curves on figures~\ref{fig:ClsContributions} and \ref{fig:XS}) get constant for smaller and smaller values of the DM mass, once the mediator mass is fixed (also in \SM scenarios when the DM mass is smaller than 100 GeV). This behaviour originates from the interplay between the functional dependence of the $\lambda$ coupling on $M_X$ and $M_Y$ in the different models,
\begin{equation}\begin{split}
  \SM: \lambda &\propto \frac{M_Y^2}{M_Y^2 - M_X^2} \;, \\
  \FS: \lambda &\propto \frac{M_Y^2}{M_Y^2 - M_X^2} \;, \\
  \FV: \lambda &\propto \frac{M_X M_Y^2}{\sqrt{2M_X^6-3M_X^4M_Y^2+M_Y^6}} \;,
\end{split}\end{equation}
and that of the associated $YY_t$ matrix elements squared. After ignoring constant numerical factors (including the fixed value of $\Gamma_Y/M_Y$), the latter are given, for $uu$-initiated and $u\bar u$-initiated processes, by
\begin{equation}\begin{split}
\hspace*{-.25cm}&\SM: 
   \left\{\begin{array}{l} 
     \mathcal M^2_{uu}  \propto \lambda^4 \frac{t u - M_Y^2}{(t-M_X^2)^2} \\
     \mathcal M^2_{u\bar u} \propto \lambda^4 \frac{ s M_X^2}{(t-M_X^2)^2}
   \end{array}\right.\ ,\\
\hspace*{-.25cm}&\FS: 
    \mathcal M^2_{uu} = \mathcal M^2_{u\bar u} \propto \lambda^4 \frac{ (t-M_Y^2)^2}{(t-M_X^2)^2}  \ ,\\
\hspace*{-.25cm}&\FV:
  \left\{\begin{array}{l}
    \mathcal M^2_{uu} \propto \lambda^4 \frac{ \left[2M_X^2(t-M_Y^2)+M_Y^2(t+2u-3M_Y^2)\right]^2}{M_X^4(t-M_X^2)^2} \\
    \mathcal M^2_{u\bar u} \propto \lambda^4 \frac{\left[2M_X^2(M_Y^2-t)+M_Y^2(s-u+M_Y^2)\right]^2}{M_X^4(t-M_X^2)^2}
  \end{array}\right.\hspace*{-.15cm},
\end{split}\label{eq:YYt}\end{equation}
where $\mathcal M^2_{uu}$ and $\mathcal M^2_{u\bar u}$ refer to the amplitude squared relevant for the $uu\to YY$ and $u\bar u \to YY^*$ processes respectively. Whereas  $\mathcal M^2_{uu} = \mathcal M^2_{\bar u\bar u}$, $\bar u\bar u$-initiated processes are less relevant as relatively suppressed by parton densities. They are thus ignored in the current discussion. In scenarios with a fermionic mediator (\FS and \FV), none of the two $YY_t$ processes $uu\to YY$ and $u \bar u\to YY^*$ depends on $M_X$ in the limit of small $M_X$. On the contrary, in the scalar mediator scenario only the amplitude associated with the $uu$-initiated process is independent of $M_X$ in the same limit, the other amplitude decreasing with smaller and smaller $M_X$ values. This then explains why in \SM models the limits become independent of $M_X$ only when the DM mass is not too large, in contrast to other scenarios.

\subsubsection{Scenarios with fixed $\lambda$ coupling}\label{sec:collider_lam}

\begin{figure*}
\includegraphics[width=.32\textwidth]{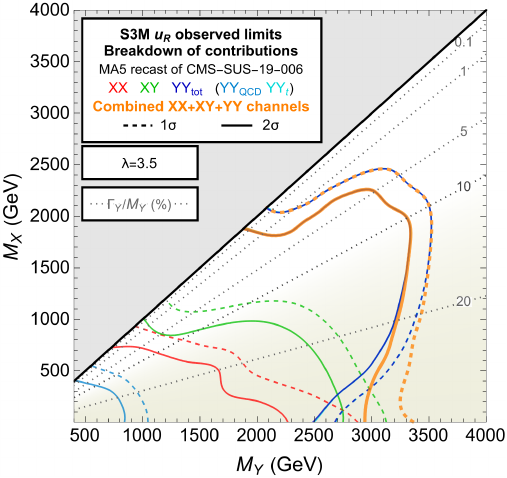}
\includegraphics[width=.32\textwidth]{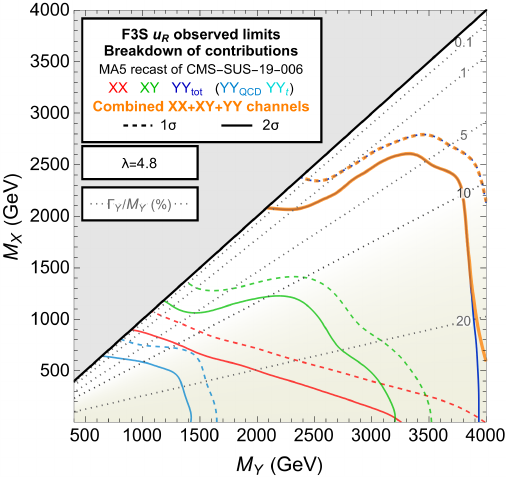}
\includegraphics[width=.32\textwidth]{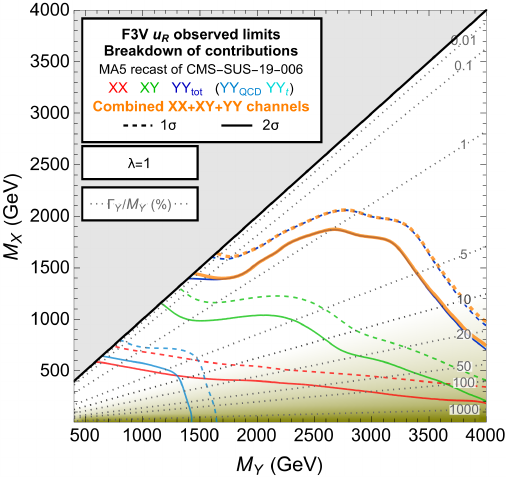}
\caption{\label{fig:ClsContributionsCoup}Same as figure~\ref{fig:ClsContributions}, but for scenarios in which the $\lambda$ coupling is fixed to $\lambda=3.5$ (\SM, left), $\lambda=4.8$ (\FS, centre) and $\lambda=1$ (\FV, right). The area with yellow gradient indicates when the $\Gamma_Y/M_Y$ ratio becomes so large that a treatment involving the narrow-width approximation becomes less and less valid.}
\end{figure*}

In this section we present our results in an alternative fashion. Instead of fixing the mediator's width-over-mass ratio, we fix the coupling $\lambda$ for all points in the $(M_Y, M_X)$ plane to a common value. It is important to keep in mind that results obtained under this assumption have to be interpreted carefully: a non-constant $\Gamma_Y/M_Y$ ratio means that the narrow-width approximation might not be valid in some regions of the parameter space. However, in our simulations mediator production and decay are factorised, which can only be achieved when the mediator width is small enough relative to the mediator mass. For this reason, we have considered different coupling values for the different scenarios such that in large part of the mass-mass planes shown the NWA is ensured. We (arbitrarily) adopt $\lambda=3.5$ for \SM models, $\lambda=4.8$ for \FS models, and $\lambda=1$ for \FV models, those large values being nevertheless motivated by the astrophysical and cosmological bounds discussed in~section~\ref{sec:cosmo}. The bounds obtained through the reinterpretation of the results of the CMS-SUS-19-006 analysis~\cite{CMS:2019zmd} are displayed in figure~\ref{fig:ClsContributionsCoup} for the \SM (left), \FS (centre) and \FV (right) scenarios. We additionally indicate through a yellow gradient the regions of the parameter space in which the mediator width-to-mass ratio is larger than 10\%.

As in~section~\ref{sec:collider_GoverM}, in the NWA region the bounds are entirely driven by the $YY_t$ channel for all scenarios. An interplay between the $YY_t$ and $XY$ modes seems to emerge for \SM, but as the width of the mediator is above 20\% of its mass, results in this area may be inaccurate. In the latter case and for the adopted coupling value of $\lambda=3.5$ (left panel of the figure), mediator masses ranging up to $M_Y\simeq 3$~TeV are excluded at 95\% confidence level for DM masses below $M_X\simeq 2$~TeV, the bounds vanishing otherwise. Similar exclusions are found for \FS models and $\lambda=4.8$, the exclusion contour boundaries being this time given by $M_Y \simeq 4$~TeV and $M_X\simeq 2.5$~TeV. For \FV scenarios and $\lambda=1$, mediator masses higher than 4~TeV could in principle be reached. However, each of these setups is ill-defined as they would correspond to $\Gamma_Y/m_Y > 20\%$. Conversely, for $M_Y\lesssim 4$~TeV DM masses smaller than 1--1.5~TeV are found excluded at 95\% confidence level.

The above findings exhibit a remarkable difference between the \FV models and the other scenarios. In the \FV case, scenarios featuring a small $M_X$ value are excluded even for extremely large values of $M_Y$. This difference can be explained by the dependence on $M_X$ of the squared matrix elements associated with the $YY_t$ processes shown in \eqref{eq:YYt}. Only for \FV the amplitude squared increases for decreasing $M_X$ values. All others processes include a component independent of $M_X$ that becomes at some point dominant for decreasing DM masses. Moreover, for scenarios in which $M_X$ is small, the $XX$ process could also be relevant. On the one hand, we get a phase-space enhancement for small $M_X$, and on the other hand the amplitude squared satisfies, in the different models,
\begin{equation} \begin{split}
&\SM: \mathcal M^2_{XX} \propto \lambda^4 \frac{(M_X^2 - t)^2}{(t-M_Y^2)^2}\ ,\\
&\FS: \mathcal M^2_{XX} \propto \lambda^4 \frac{t u -M_X^4}{(t-M_Y^2)^2}   \ ,\\
&\FV: \mathcal M^2_{XX} \propto \lambda^4 \frac{M_Y^2}{M_X^4} \\ 
   &\hspace*{.5cm} \times \frac{4 M_X^6 - 4 M_X^4 (s+2t) +4 M_X^2 t(s+t) - s t^2}{(t-M_Y^2)^2}  \ .
\end{split}\label{eq:XX}\end{equation}

The enhancement of the $XX$ production cross section in \FV scenarios when $\lambda$ is constant, despite the low sensitivity of the CMS-SUS-19-006 analysis~\cite{CMS:2019zmd} for this channel, is sufficient to lead to an exclusion bound when DM is light. While the contribution is still subdominant relative to that of $YY_t$, the $XX$ channel is not affected by the adopted NWA assumption. Therefore, signal regions targeting specifically DM pair production (together with a reduced hadronic activity) and sensitive to it could become a discriminating handle to characterise different DM scenarios.

\section{Complementarity between astrophysics, cosmology and LHC constraints}
\label{sec:cosmoLHCbounds}

\begin{figure*}
\includegraphics[width=.32\textwidth]{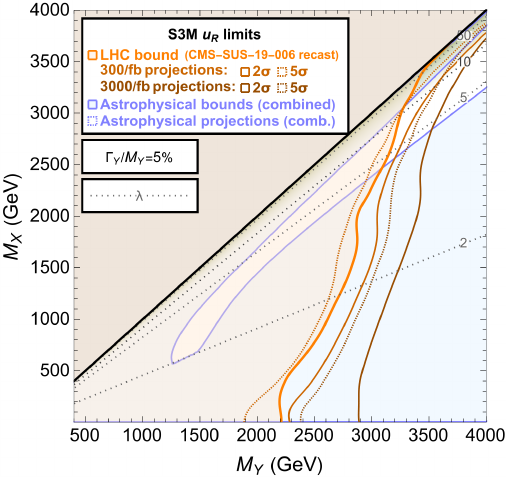}
\includegraphics[width=.32\textwidth]{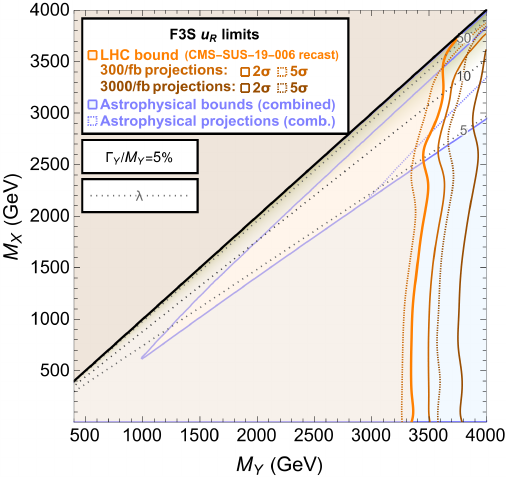}
\includegraphics[width=.32\textwidth]{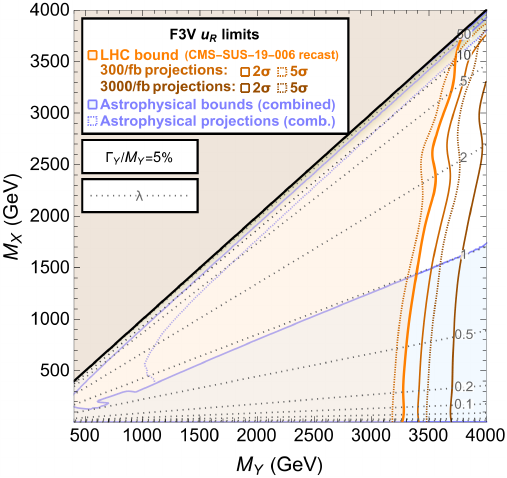}
\includegraphics[width=.32\textwidth]{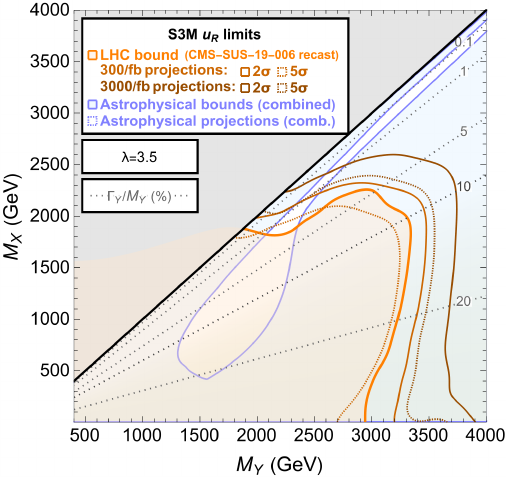}
\includegraphics[width=.32\textwidth]{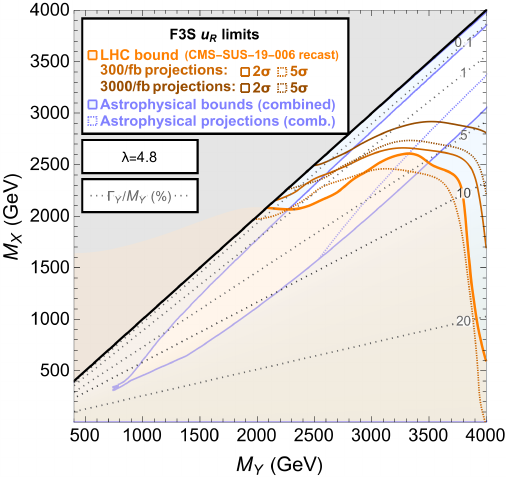}
\includegraphics[width=.32\textwidth]{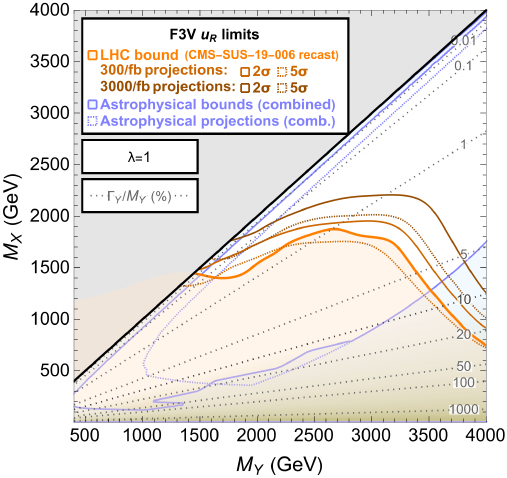}
\caption{\label{fig:combinationCosmoLHC} Combination of the cosmological, astrophysical and collider constraints discussed in this study for the three real DM scenarios described in section~\ref{sec:theory}. We consider tle \SM (left), \FS (centre) and \FV (right) classes of models in which either the new physics couplings value is determined by fixing the mediator width-over-mass ratio $\Gamma_Y/M_Y$ to 0.05 (top row), or it is fixed to a specific value (bottom row). Projections for future astrophysical and collider bounds are also provided, the LHC projections being achieved under the assumption that systematics errors $\Delta_{\rm bkg}$ scale with the luminosity $\mathcal{L}$ as $\Delta_{\rm bkg}/\sqrt{\mathcal{L}}={\rm constant}$.}
\end{figure*}

In figure~\ref{fig:combinationCosmoLHC} we highlight the complementarity exhibited by all cosmological, astrophysical and collider constraints explored, both for $t$-channel DM scenarios in which the mediator width-over-mass ratio is fixed and for scenarios in which the new physics coupling is instead set to a specific value. We consider \SM (left), \FS (centre) and \FV (right) models with a self-conjugate DM state (complex DM options being excluded by cosmology and astrophysics, as discussed in section~\ref{sec:cosmo}). Scenarios excluded by cosmological and astrophysics constraints are shown in blue, whereas scenarios excluded by DM searches at colliders are shown in orange. We recall that for all the results presented in this section, we focus on scenarios of class (ii) in which DM is allowed to be under-abundant. The lower boundary of the regions excluded by cosmology and astrophysics (blue areas) therefore corresponds to scenarios featuring a relic density in agreement with Planck data, whereas the non-blue areas (and the allowed white areas in particular) always correspond to scenarios with under-abundant DM, when it is assumed to only consist of the $X$ state. Multi-component DM must therefore be invoked to restore agreement with data in such new physics setups.

When $\Gamma_Y/M_Y=0.05$ (top row of figure~\ref{fig:combinationCosmoLHC}) the regions of the parameter spaces allowed by all constraints always correspond to configurations featuring mediator masses $M_Y$ larger than 3--3.5~TeV, while the lower allowed values for the DM mass $M_X$ differ from case to case. They range from 1.5 TeV for \FV models to 2.5 TeV for \SM models. On the other hand, as in section~\ref{sec:collider_lam} we study \SM, \FS and \FV scenarios with a specific coupling value $\lambda=3.5$, 4.8 and 1 respectively (bottom row of figure~\ref{fig:combinationCosmoLHC}), those values being the lowest ones leading to parameter configurations not excluded by cosmological and astrophysical bounds (see section~\ref{sec:cosmo}). The combination with collider constraints further imposes a lower limit on the mediator mass ranging from 1.5~TeV for \FV models to 2~TeV in \SM and \FS scenarios. The lower bound on the DM mass is instead still mostly driven by cosmology and astrophysics, and it lies in the 1.5--2~TeV regime.

Figure~\ref{fig:combinationCosmoLHC} also includes projections for the current astrophysical and LHC bounds in light of future data. The impact of future astrophysical experiments is found to differ from one scenario to another. In the \SM class of models (first column of the figure), the expected reach of the PICO-500 experiment is fully complementary to the constraints originating from the relic density so that the whole parameter space (both for scenarios with a fixed width-over-mass ratio and those with a fixed new physics coupling) could be potentially excluded. \FV models (right column of the figure) exhibit a similar behaviour so that the results expected from the PICO-500 experiment will significantly improve the astrophysical coverage of the model's parameter space. However, a large amount of configurations will this time be left unexplored. On the contrary, in the \FS class of models (central column of the figure) will mostly resist to future DD and ID searches for DM. Future experiments are indeed only expected to be sensitive to scenarios located in a small additional part of the currently allowed parameter space, leaving many options uncovered and open for further exploration, {\it e.g.}, at colliders.

To assess the future sensitivity of the LHC to the models studied, we determine projections for the two nominal luminosities $\mathcal{L}=300$~fb$^{-1}$ and 3000~fb$^{-1}$, corresponding to the end of the third operation run of the LHC and to its high-luminosity (HL-LHC) phase, respectively. Bounds are computed under the optimistic assumption that the systematic uncertainties on the background $\Delta_{\rm bkg}$ will be reduced and scale as $\Delta_{\rm bkg}/\sqrt{\mathcal{L}}={\rm constant}$. Projected discovery reaches (for a significance of 5$\sigma$) are also included, demonstrating that the expected improvement is sizeable, especially for what concerns the HL-LHC phase. The gain in parameter space coverage hence virtuously complements the expected reach of future astrophysical experiments for all scenarios explored, and only benchmark setups with mediator and dark matter masses lying deep in the TeV regime are expected to survive.

\section{Conclusions}\label{sec:concl}
In this work, we explore $t$-channel simplified models of dark matter in which the Standard Model is extended by one DM state $X$ and one coloured mediator state $Y$. Both new fields are taken to be odd under a new $\mathbb{Z}_2$ parity, the SM fields being instead enforced to be even, so that the theory only features a single new physics coupling vector in the flavour space. For simplicity, we consider models in which the dark matter solely couples to the right-handed up quark, and we additionally focus on different possibilities for the spin of the new particles and the self-conjugate properties of the dark matter. We hence study six cases with a tri-dimensional parameter space defined by the mass of the dark matter $M_X$, the mass of the mediator $M_Y$, and the new physics coupling $\lambda$. The dark matter is taken to be either a scalar field, a fermion field or a vector field, and it could be self-conjugate or not. The mediator is consequently either a scalar particle (for fermionic DM cases) or a fermion (for bosonic DM cases).

In our study, we investigate which configurations of the free parameters of the six models are compliant with constraints originating from the relic density, astrophysical probes of dark matter and searches for DM at colliders. Our results reveal a virtuous complementarity between the different probes, leading to an excellent coverage of the six parameter spaces that has the potential to be further improved in the near future. 

Notably, requiring that dark matter is not over-abundant and imposing constraints from direct and indirect DM searches suffice to exclude the entire parameter space of all models featuring complex DM, except for a region of small mass splitting between the dark matter particle and the mediator. In this region, the measured relic density can be explained for very weak couplings $\lambda$ for which dark matter genesis proceeds via conversion-driven freeze-out, predicting a long-lived mediator. In the self-conjugate DM cases studied, several scenarios with heavy new physics particles are still allowed, together with a few exceptions at lower masses. Collider constraints, however, push the bounds deep into the TeV regime. This improves our previous results~\cite{Arina:2020tuw} by several hundreds of GeV in the parameter region allowed also by astrophysical observations, and originates in particular from a proper modelling of the associated signals including the highly relevant contribution from same-sign mediator pair production (that has still not been considered experimentally so far). Mediator lower mass limits are found to be of 3--4~TeV for various hypersurfaces in the tri-dimensional model parameter spaces. These hypersurfaces are defined by either setting the mediator width-over-mass ratio to some value (5\% in the cases studied), or by fixing the coupling $\lambda$ itself directly to a specific value (that we choose to be 3.5, 4.8 and 1 in the \SM, \FS and \FV classes of models, in agreement with cosmological and astrophysical exclusions and leading to a narrow-width mediator for most of the mass values which can be explored at the LHC). 

Our findings further show that future direct detection experiments and LHC searches with a luminosity of 3000~fb$^{-1}$ have the power to entirely exclude the possibility of Majorana dark matter in  the prompt regime, and to very strongly restrict bosonic dark matter options. While $t$-channel simplified models of dark matter are still interesting benchmarks in the context of DM searches today, future data is thus guaranteed to provide further insights into the models and to maximise their potential as representative scenarios for large classes of UV-complete setups.

\section*{Acknowledgements} 
We are grateful to J.~Salko for discussions in the earlier phase of this work, as well as to F. Benoit, L. Munoz Aillaud and G. Tortarolo for having brought to our attention specific issues with \mg\ simulations, {\sc MadSpin} and the models considered. We thank M.~Garny for useful comments, and furthermore acknowledge the use of the IRIDIS HPC Facility at the University of Southampton.

This work has been supported by the French ANR (grant ANR-21-CE31-0013, `DMwithLLPatLHC'), the F.R.S.-FNRS under the “Excellence of Science” EOS be.h project no.~30820817, the European Research Council under the European Union’s Horizon 2020 research and innovation Programme (grant agreement n.950246), and by the Deutsche Forschungsgemeinschaft (DFG, German Research Foundation) under grant 396021762 - TRR 257. J.H.~acknowledges support by the Alexander von Humboldt foundation via the Feodor Lynen Research Fellowship for Experienced Researchers and the Feodor Lynen Return Fellowship. LP's work is (partially) supported by the ICSC -- Centro Nazionale di Ricerca in High Performance Computing, Big Data and Quantum Computing, funded by the European Union (NextGenerationEU).

\appendix

\section{Annihilation cross sections}
\label{app:cross_sections}

In this appendix we provide analytical formulas for the DM annihilation cross sections in the various models 
studied in this work. The class of models \SD, \FV and \FW are all characterised by a cross section dominated by $s$-wave annihilations in the $u \bar{u}$ final state. The associated expressions can be written in a compact manner in terms of the DM mass $M_X$, the new physics coupling $\lambda$ and the mass ratio between the dark matter and the mediator $r = M_X / M_Y$,
\begin{equation}\begin{split}
    \langle\sigma v\rangle_{\SD} =&\ \frac{3 \lambda^4}{64 \pi M_X^2 (1 + r^2)^2} \, ,\\
    \langle\sigma v\rangle_{\FV} =&\ \frac{2 \lambda^4}{3 \pi M_X^2 (1 + r^2)^2} \, ,\\
    \langle\sigma v\rangle_{\FW} =&\ \frac{\lambda^4}{12 \pi M_X^2 (1 + r^2)^2} \, .
\end{split}\end{equation}
For the rest of the models explored (\SM , \FS and \FC), DM annihilation into quarks is characterised by a helicity suppression which induces a velocity dependence of the cross section. This suppression can however be lifted by considering additional gluon or photon emissions ($XX\to u\bar u \gamma$ or $u\bar u g$). In addition, loop-induced annihilations into pairs of photons or gluons could now play a role too. The $s$-wave leading contributions to the annihilation cross sections are therefore given by
\begin{equation}\begin{split}
    \langle\sigma v\rangle_{\SM}^{\gamma \gamma/ gg} =&\ K_i \frac{\alpha_i^2 \lambda^4}{144 \pi^3 M_X^2} \mathcal{I}(r)^2 \, ,\\
    \langle\sigma v\rangle_{\FX}^{\gamma \gamma/ gg} =&\ K_i \frac{\alpha_i^2 \lambda^4}{18 \pi^3 M_X^2 F_X} \mathcal{A}(r)^2 \, ,\\
    \langle\sigma v\rangle_{\SM}^{u \bar{u} (\gamma/g)} =&\ K^\prime_i \frac{\alpha_i \lambda^4}{48 \pi^2 M_X^2} f(r) \, ,\\
    \langle\sigma v\rangle_{\FX}^{u \bar{u} (\gamma/g)} =&\ K^\prime_i \frac{ \alpha_i \lambda^4}{6 \pi^2 M_X^2 F_X} f(r) \, ,
\end{split}\end{equation}
for \texttt{X} $=$ \texttt{S} or \texttt{C}, and $F_X = 1$ or 8 respectively. In those expressions the index $i$ refers to a final state comprising photons ($\gamma$) or gluons ($g$) such that $K_\gamma = 1$, $K_g=9/8$, $K_\gamma^\prime = 1$, $K_g^\prime = 3$, and $\alpha_\gamma$ and $\alpha_g$ respectively referring to the electromagnetic and strong coupling constants $\alpha$ and $\alpha_s$.  The functions depending on $r$ are given by
\begin{equation}
    \mathcal{I}(r) =  \int_{0}^{1} \frac{{\rm d} x}{x} \log \left|\frac{-x^{2}+\left(1-r^{2}\right) x+r^{2}}{x^{2}+\left(-1-r^{2}\right) x+r^{2}}\right| \, ,\nonumber 
\end{equation}
\begin{equation}\begin{split}
    \mathcal{A}(r) =&\ 2 \!+\! \Li\left[1\!-\!\frac{1}{r^{2}}\right] \!-\! \Li\left[1\!+\!\frac{1}{r^{2}}\right] \!-\! 2 r^{2} \arcsin ^{2}\frac{1}{r} \, , \\
   f(r) =&\ \Big(r^{2}+1\Big) \left(\frac{\pi^{2}}{6} \!-\! \log \left[\frac{r^{2}\!+\!1}{2 r^{2}}\right]^{2} \!-\! 2 \Li\left[\frac{r^{2}+1}{2 r^{2}}\right]\right) \\ &\quad
   + \! \frac{4 r^{2}\!+\!3}{r^{2}\!+\!1} \!+\! \frac{4 r^{4}\!-\!3 r^{2}\!-\!1}{2 r^{2}} \log \frac{r^{2}-1}{r^{2}+1}\, .
\end{split}\end{equation}

\newpage

\bibliographystyle{JHEP}
\bibliography{biblio}

\end{document}